\documentclass[aps,pra,reprint,twocolumn,superscriptaddress,floatfix,longbibliography]{revtex4-1}
\usepackage{graphicx,amsmath,amsfonts,amssymb,amsthm,xr}
\usepackage{epsfig,amsmath,amssymb,color,dsfont,upgreek,physics,bm}
\usepackage{mathrsfs}
\usepackage{mathtools}
\usepackage{bbold}
\usepackage{comment}
\usepackage{qcircuit}
\usepackage{float}

\usepackage{braket}
\usepackage[bookmarks=true,colorlinks,linkcolor=OrangeRed,urlcolor=NavyBlue,citecolor=RoyalBlue]{hyperref}
\usepackage[dvipsnames]{xcolor}
\usepackage{orcidlink}

\graphicspath{{./Figures/}}

\begin{document}
	\title{Robust finite-temperature many-body scarring on a quantum computer}
	\author{Jean-Yves Desaules${}^{\orcidlink{0000-0002-3749-6375}}$}
	\affiliation{School of Physics and Astronomy, University of Leeds, Leeds LS2 9JT, UK}
	\author{Erik J.~Gustafson$^{\orcidlink{0000-0001-7217-5692}}$}
	\affiliation{Quantum Artificial Intelligence Laboratory (QuAIL), NASA Ames Research Center, Moffett Field, CA, 94035, USA}
	\affiliation{USRA Research Institute for Advanced Computer Science (RIACS), Mountain View, CA, 94043, USA}
	\author{Andy C.~Y.~Li${}^{\orcidlink{0000-0003-4542-3739}}$}
	\affiliation{Fermi National Accelerator Laboratory, Batavia, Illinois, 60510, USA}
	\author{Zlatko Papi\'c${}^{\orcidlink{0000-0002-8451-2235}}$}
	\affiliation{School of Physics and Astronomy, University of Leeds, Leeds LS2 9JT, UK}
	\author{Jad C.~Halimeh${}^{\orcidlink{0000-0002-0659-7990}}$}
	\affiliation{Department of Physics and Arnold Sommerfeld Center for Theoretical Physics (ASC), Ludwig-Maximilians-Universit\"at M\"unchen, Theresienstra\ss e 37, D-80333 M\"unchen, Germany}
	\affiliation{Munich Center for Quantum Science and Technology (MCQST), Schellingstra\ss e 4, D-80799 M\"unchen, Germany}
	
	\begin{abstract}
		Mechanisms for suppressing thermalization in disorder-free many-body systems, such as Hilbert space fragmentation and quantum many-body scars, have recently attracted much interest in foundations of quantum statistical physics and potential quantum information processing applications. However,  their sensitivity to realistic effects such as finite temperature remains largely unexplored. Here, we have utilized IBM's Kolkata quantum processor to demonstrate an unexpected robustness of quantum many-body scars at finite temperatures when the system is prepared in a thermal Gibbs ensemble. We identify such robustness in the PXP model, which describes quantum many-body scars in experimental systems of Rydberg atom arrays and ultracold atoms in tilted Bose--Hubbard optical lattices. By contrast, other theoretical models which host exact quantum many-body scars are found to lack such robustness, and their scarring properties quickly decay with temperature. 
		Our study sheds light on the important differences between scarred models in terms of their algebraic structures, which impacts their resilience to finite temperature.
	\end{abstract}
	
	\date{\today}
	\maketitle
	
	\section{Introduction} 
	
	The development of programmable Rydberg atom arrays~\cite{Bernien2017} (see also the review~\cite{Browaeys_review}) has ushered in an era of experimental explorations of a weak breakdown of thermalization, now commonly referred to as quantum many-body scars (QMBSs)~\cite{Serbyn2021Review, MoudgalyaReview, ChandranReview}. In QMBS systems, only a small (typically vanishing in system size) fraction of eigenstates violate the Eigenstate Thermalization Hypothesis (ETH)~\cite{Deutsch1991,Srednicki1994}, while the rest of the many-body spectrum is chaotic. Such systems exhibit thermalizing dynamics from most initial states, however their dynamics can be strikingly regular for a small set of special initial conditions, as indeed observed in experiments~\cite{Bernien2017, Bluvstein2021, Su2022, Zhang2023}. This ``intermediate'' behavior between chaos and integrability has attracted attention in the context of controlling quantum dynamics~\cite{Bluvstein2021} and entanglement~\cite{Dong2023} in complex systems and using such systems for quantum-enhanced metrology~\cite{Dooley2021, DesaulesQFI, Dooley2022}.
	
	Given the strong sensitivity of scarred dynamics on the initial state, in this work we address a natural question for experiments and applications of QMBSs: how sensitive is scarring to finite temperature $T$? For example, imperfections in state preparation -- due to finite temperature -- could strongly impact the subsequent dynamics. In a scenario commonly studied in the literature,  an initial state of interest, $\ket{\psi_0}$, is prepared as the ground state of a simple preparation Hamiltonian $\hat{H}_\mathrm{i}$. The system is then quenched by rapidly changing the parameters so that the dynamics is now governed by a final Hamiltonian $\hat{H}_\mathrm{f}$, for which our prepared state is no longer necessarily close to the ground state. Here, we will consider the case where, instead of the ground state, the Gibbs state of $\hat{H}_\mathrm{i}$ at temperature $T$ is obtained as a result of preparation. 
	
	For the so-called PXP model~\cite{FendleySachdev,Lesanovsky2012,Turner2018} -- the effective model of Rydberg atom arrays mentioned above -- we find that the finite-$T$  preparation scheme still results in remarkably robust QMBS signatures, even at high temperatures. We present evidence for this based on both large-scale classical simulations as well as quantum simulation of finite-$T$ quenches on the IBM quantum computer. Surprisingly, for other models where QMBS states obey exact algebraic relations, such as the spin-$1$ XY magnet~\cite{Iadecola2019_2}, we find opposite behavior: signatures of QMBS decay fast with temperature.  Our results establish the robustness of QMBSs at finite temperature in the PXP model, and show they can be harnessed on existing quantum hardware. Moreover, they highlight the fine differences between QMBS models depending on the nature of the underlying scarring mechanism and the algebraic structure of their QMBS subspaces.
	
	The remainder of this paper is organized as follows. In Sec.~\ref{sec:protocol} we introduce our finite-temperature quench protocol and several diagnostics, in particular the inteferometric Loschmidt echo $\mathcal{F}(t)$, that we will employ to characterize thermalization and scar dynamics. In Sec.~\ref{app:XY}, as a warmup, we discuss the spin-1 XY model~\cite{Iadecola2019_2}, which exhibits exact QMBS states that are fully decoupled from the thermal bulk of the spectrum. This analytically-tractable example will allow us to establish intuition about the expected behavior of the Loschmidt echo after a finite-$T$ quench. We will show that the echo can be accurately modeled by assuming the dominant contribution comes from the ground state. Thus, in the spin-1 XY model the scar dynamics is sensitive to preparing the system in the ground state of the pre-quench Hamiltonian. By contrast, in Sec.~\ref{sec:PXP} we will study the same finite-$T$ quench protocol for the PXP model and show that the simple ground-state approximation of the echo breaks down and the post-quench dynamics exhibits \emph{more pronounced} scarring signatures than naively expected. These results are furthermore verified using the IBM quantum computer in Sec.~\ref{sec:IBM}. We conclude with a discussion of these findings in Sec.~\ref{sec:conc}, while Appendices contain further details about the Loschmidt echo and its low-temperature approximation, different preparation protocols, results for a deformed PXP model with nearly exact QMBSs, and details of the quantum algorithm executed on IBM hardware.
	
	\section{\texorpdfstring{Finite-$T$}{Finite-T} quench protocol and diagnostics of thermalization}\label{sec:protocol}
	
	To probe the effect of temperature, we prepare the system in a thermal Gibbs state at a given inverse temperature $\beta=1/T$ using some pre-quench Hamiltonian $\hat{H}_\mathrm{i}$. We then perform a quench by evolving the state with the quench Hamiltonian $\hat{H}_\mathrm{f}$. The two Hamiltonians are related, as $\hat{H}_\mathrm{i}$ must have the reviving state of $\hat{H}_\mathrm{f}$ as its ground state. 
	In addition, we also ensure that our initial state is always essentially at infinite temperature with respect to $\hat{H}_\mathrm{f}$ such that any ergodicity breaking is caused by QMBSs and not by proximity to the ground state of $\hat{H}_\mathrm{f}$, see Appendix \ref{app:temp_fin}.
	
	We will denote the eigenstates of $\hat{H}_\mathrm{i}$ by $\ket{E_n}$ and their corresponding eigenenergies $E_n$, assuming they are sorted in increasing order, $E_{n+1} \geq E_n$. The initial state is then the mixed state  
	\begin{equation}\label{eq:rho}
		\hat{\rho}(\beta)=\frac{e^{-\beta \hat{H}_\mathrm{i}}}{Z}=\frac{1}{Z}\sum_n e^{-\beta E_n}\ketbra{E_n},
	\end{equation}
	with $Z=\sum_n e^{-\beta E_n}$ the partition function of $\hat{H}_\mathrm{i}$. For simplicity, we will always add a constant diagonal contribution to $\hat{H}_\mathrm{i}$ to ensure that the ground state has energy $E_0=0$, which has no impact on the physics but simplifies the calculations. 
	
	At time $t=0$, we quench the system with the Hamiltonian $\hat{H}_\mathrm{f}$, generally distinct from $\hat{H}_\mathrm{i}$, and let it evolve freely as a closed system. We characterize the dynamics by the interferometric Loschmidt echo 
	\begin{align}\label{eq:LE}
		\mathcal{F}(t)=\big\lvert{\Tr}\big\{e^{-i\hat{H}_\mathrm{f}t}\hat{\rho}\big\}\big\rvert^2,
	\end{align}
	which is a suitable generalization of the more familiar return fidelity, to which it reduces in the case of a pure state. If $\hat{H}_\mathrm{f}$ obeys the ETH and $\hat{\rho}$ is close to an infinite-temperature state with respect to $\hat{H}_\mathrm{f}$, then we expect $\cal F$ to quickly approach $1/\mathcal{D}$, with $\mathcal{D}$ the Hilbert space dimension. On the other hand, after a quench from a scarred initial state, we expect $\mathcal{F}(t)$ to return to an $\mathcal{O}(1)$ value after some number of cycles with period $\tau$. As such, the main quantity we will investigate is $\mathcal{F}_k$, which is the maximum of $\mathcal{F}(t)$ in the vicinity of $t=k\tau$. When performing system-size scaling to the thermodynamic limit, we will also use the fidelity density, $f=\ln\left(\mathcal{F}\right)/N$, which is an intensive quantity. We will use the same notation of $f_k$ to denote $\ln\left(\mathcal{F}_k\right)/N$.
	
	In order to detect anomalous responses at finite temperature, it is important to have some intuition about the behavior of $\mathcal{F}_k$. The simplest conjecture one could make is that only the ground state contributes to the revivals. We then expect $\mathcal{F}_k$ to behave as $\mathcal{F}_k=\mathcal{F}^\infty_k/Z^2$ away from $\beta=0$, where $\mathcal{F}^\infty_k$ is the value at $\beta=\infty$ which is equal to $1$ in the case of ``perfect'' scarring. While our focus is mostly on the regime of large $\beta$ where signatures of scarring have a chance of being measured in experiment, we can refine our approximation to also make predictions as $\beta$ gets closer to 0. The full derivation can be found in Appendix~\ref{app:LowT}, with the final result
	\begin{equation}\label{eq:Fth}
		\mathcal{F}_k= \frac{\mathcal{F}^\infty_k}{Z^2}
		+\frac{\left(1-\frac{1}{Z}\right)^2}{\mathcal{D}}.
	\end{equation}
	Due to several approximations involved in the derivation,  this formula is rather crude for low $\beta$ and is only meant to give an idea of the magnitude of $\mathcal{F}_k$ in that regime. However, we will show that in the spin-1 XY model our prediction agrees well with numerical results over the full range of $\beta$.
	
	As a second diagnostic, we turn towards observables. In particular, we study the density of $\hat{H}_\mathrm{i}$ defined as $\hat{h}=\hat{H}_\mathrm{i}/N$. The expectation value of $\hat{h}$ at time $t$ is given by $\braket{\hat{h}}_\beta(t) \equiv \Tr(\hat{\rho}_\beta(t)\hat{h})$, where we use angular brackets $\langle \ldots \rangle_\beta$ to denote an expectation value in the Gibbs state at inverse temperature $\beta$. As our focus is on initial states at infinite temperature with respect to the quench Hamiltonian $\hat{H}_\mathrm{f}$, we are interested in the deviation of $\hat{h}$ from its infinite-$T$ expectation value:
	\begin{eqnarray}
		\delta \hat{h}=(\hat{h}-\langle \hat{h} \rangle_{\beta=0})/\langle \hat{h} \rangle_{\beta=0}, 
	\end{eqnarray}
	Note that $\hat{h}$ is positive semi-definite by construction as the ground state energy was set to zero. As $\hat{h}$ is not proportional to the identity, it must have strictly positive eigenvalues and thus $\langle \hat{h} \rangle_{\beta=0}$ (equal to the mean of the eigenvalues) cannot be zero, meaning that $\delta \hat{h}$ is never singular. 
	
	Similar to the Loschmidt echo, we will focus on the observable expectation value after $k$ periods. In the majority of scarred models, scarring dynamics can be viewed as state transfer between two product states. For both models that we study in this work, we additionally have that these two states are extremal eigenstates of $\hat{H}_\mathrm{i}$. As such, we expect $\braket{\hat{h}}(t)$ and $\braket{\delta \hat{h}}$ to be maximal at $t=(k+1/2)\tau$ and minimal at $t=k\tau$ with $k$ integer. To minimize clutter, we will use an abbreviation for the expectation value $\braket{\delta \hat{h}}_\beta(k\tau) \equiv \delta h_k$, keeping the inverse temperature $\beta$ implicit, and we will consider $k$ both integer and half-integer.  Analogous to Eq.~\eqref{eq:Fth}, we can derive the expected behavior in large systems to be 
	\begin{equation}\label{eq:obs_approx}
		\begin{aligned}
			\delta h_k=  \left(h_k^\infty/\braket{\hat{h}}_{\beta=0}-1\right)/Z,
		\end{aligned}
	\end{equation}
	where $h_k^\infty$ is the value at zero temperature at $t=k\tau$, which reduces to zero in the case of perfect scarring.
	The simplicity of this expression comes from the various conditions we have imposed on our initial state -- see Appendix~\ref{app:LowT}.
	
	\section{\texorpdfstring{Spin-$1$}{Spin-1} XY model}\label{app:XY}
	To test our theoretical predictions, we first apply the diagnostics introduced above to the finite-$T$ quenches of a one-dimensional (1D) spin-$1$ XY magnet~\cite{Iadecola2019_2}, in which the QMBS eigenstates can be exactly constructed. The spin-1 XY model is described by the Hamiltonian  
	\begin{align}\nonumber
		\hat{H}_\mathrm{f,XY}&=\! J\sum_{j=1}^{N-1} \left(\hat{S}^x_j\hat{S}^x_{j+1}+\hat{S}^y_j\hat{S}^y_{j+1}\right)\\\nonumber
		&+h\sum_{j=1}^N\hat{S}^z_j+D \sum_{j=1}^N\left(\hat{S}^z_i\right)^2\\\label{eq:Hf_XY}
		&+J_3\sum_{j=1}^{N-3} \left(\hat{S}^x_j\hat{S}^x_{j+3}+\hat{S}^y_j\hat{S}^y_{j+3}\right),
	\end{align}
	where $\hat{S}_j^\alpha$ are the standard spin-1 operators on site $j$. Unless specified otherwise, we will set $J=1$, $h=1$, $D=0.1$, and $J_3=0.1$ and assume open boundary conditions (OBCs). For these values of parameters, the model was shown to be non-integrable and displays chaotic level statistics~\cite{Iadecola2019_2}. At the same time, preparing the system in the initial state
	\begin{equation}\label{eq:psi0_XY}
		\ket{\psi_0}=\bigotimes_{j=1}^N \left[\frac{\ket{+1}-(-1)^j\ket{-1}}{\sqrt{2}}\right]
	\end{equation}
	was shown to give rise to perfect oscillatory dynamics~\cite{Iadecola2019_2}, revealing the existence of QMBSs. The $N+1$ special eigenstates of $\hat{H}_\mathrm{f,XY}$ have a simple form, being the superposition of all configuration with $m$ sites equal to $\ket{1}$ and $n-m$ sites equal to $\ket{-1}$. Essentially, they correspond to forming a large spin in the effective spin-1/2 subspace formed by $\{\ket{-1},\ket{1}\}$ on each site.  
	The QMBS oscillations can then be understood as the precession of this large spin, with the initial state $\ket{\psi_0}$ in Eq.~\eqref{eq:psi0_XY} having overlap only with the scarred eigenstates.  
	This motivates our choice of this model, as it admits a similar description of QMBS dynamics to the PXP model, studied in Sec.~\ref{sec:PXP} below, but with the added benefit of a closed analytic form for the QMBS eigenstates. 
	
	\begin{figure}[tb]
		\centering
		\includegraphics[width=\linewidth]{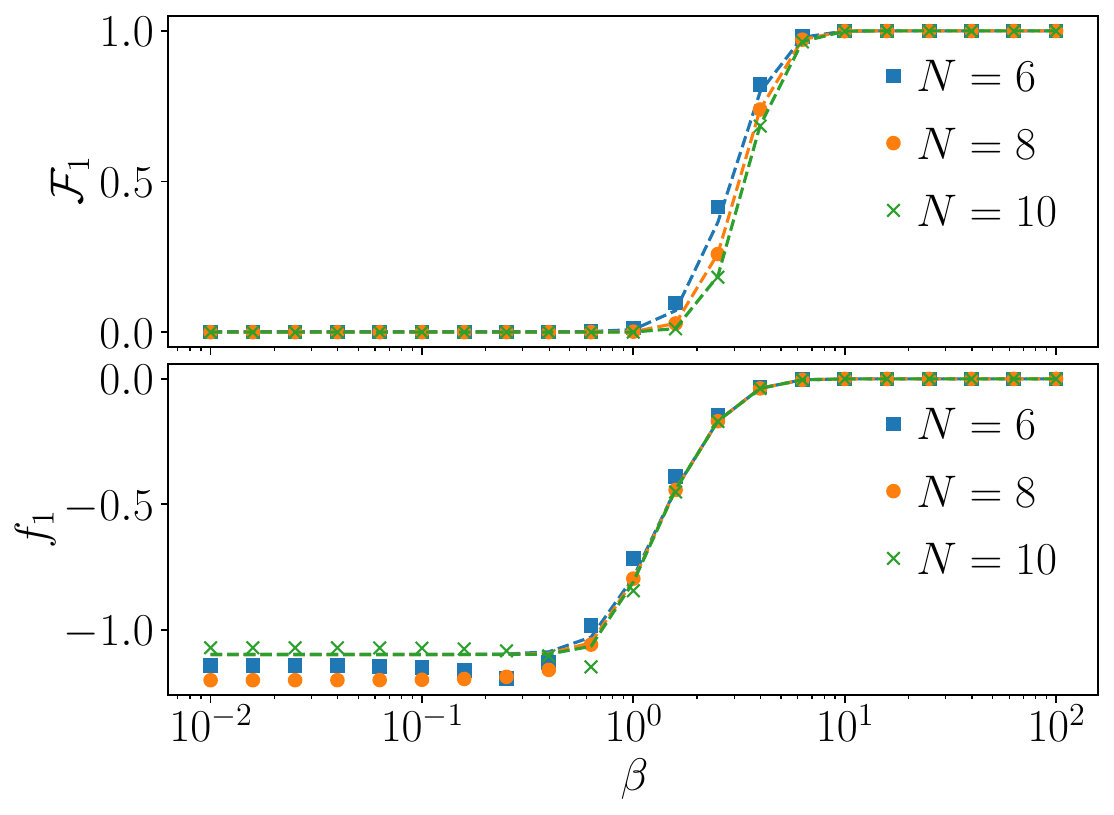}
		\caption{\small Maximum interferometric amplitude after a quench in the spin-1 XY model for various temperatures using the preparation Hamiltonian in Eq.~\eqref{eq:Hi_XY}. The dashed lines show the expected scaling in Eq.~\eqref{eq:Fth}, with the partition function given by Eq.~\eqref{eq:XYpartfunc}.
		}\label{fig:XY_fidNV}
	\end{figure}
	
	\begin{figure}[tb]
		\centering
		\includegraphics[width=\linewidth]{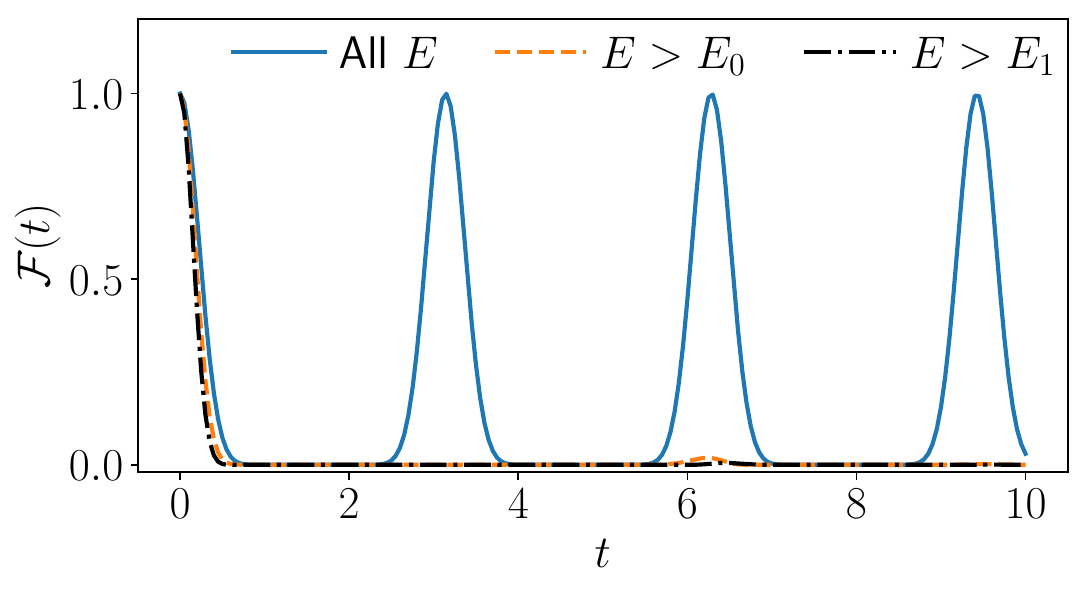}
		\caption{\small Maximum interferometric amplitude at zero temperature after a quench in the spin-$1$ XY model with $N=10$ and using $\hat{H}_\mathrm{i}$ in Eq.~\eqref{eq:Hi_XY}. Two different energy penalties on the low-energy spectrum are compared. The perfect revivals in the case without penalty (``all $E$'') are in stark contrast with their absence in cases with penalty (``$E{>}E_0$'', ``$E{>}E_1$''), showing that scarred dynamics occurs only near zero temperature. 
		}\label{fig:XY_fid_NVGS}
	\end{figure}
	
	To prepare the state in Eq.~\eqref{eq:psi0_XY}, we can use the pre-quench Hamiltonian proposed in Ref.~\cite{Iadecola2019_2}: 
	\begin{align}\label{eq:Hi_XY}
		\hat{H}_\mathrm{i,XY}= \frac{N}{2}+\sum_{j=1}^N (-1)^j \left[\left(\hat{S}_j^x\right)^2-\left(\hat{S}_j^y\right)^2\right].
	\end{align}
	Figure~\ref{fig:XY_fidNV} shows $\mathcal{F}_1$ after a quench along with its theoretical counterpart, computed using 
	\begin{equation}\label{eq:XYpartfunc}
		Z=(1+e^{-\beta}+e^{-2\beta})^N,   
	\end{equation}
	which is straightforward to derive as the Hamiltonian $\hat{H}_\mathrm{i,XY}$ is non-interacting. The agreement with the prediction is quite good, showing that states above the ground state indeed make a small contribution to the revivals.
	
	To further verify how much the excited states impact the dynamics, we investigate the scenario where an energy penalty 
	\begin{eqnarray}\label{eq:penalty}
		\hat{H}_\mathrm{i} \to \hat{H}_\mathrm{i} + V \ketbra{E_m}, \quad V\to\infty, 
	\end{eqnarray}
	is added to the preparation Hamiltonian. This essentially removes the desired eigenstate (e.g., the ground state with $m=0$) while leaving the rest of the spectrum completely untouched due to the orthogonality of eigenstates. Fig.~\ref{fig:XY_fid_NVGS} shows the  results for this modified quench, where we remove either the ground state or the first set of excited states. We see that the revivals are rapidly destroyed, except for small fluctuations that are expected to decay exponentially with system size. 
	
\begin{figure}[tb]
		\centering
		\includegraphics[width=\linewidth]{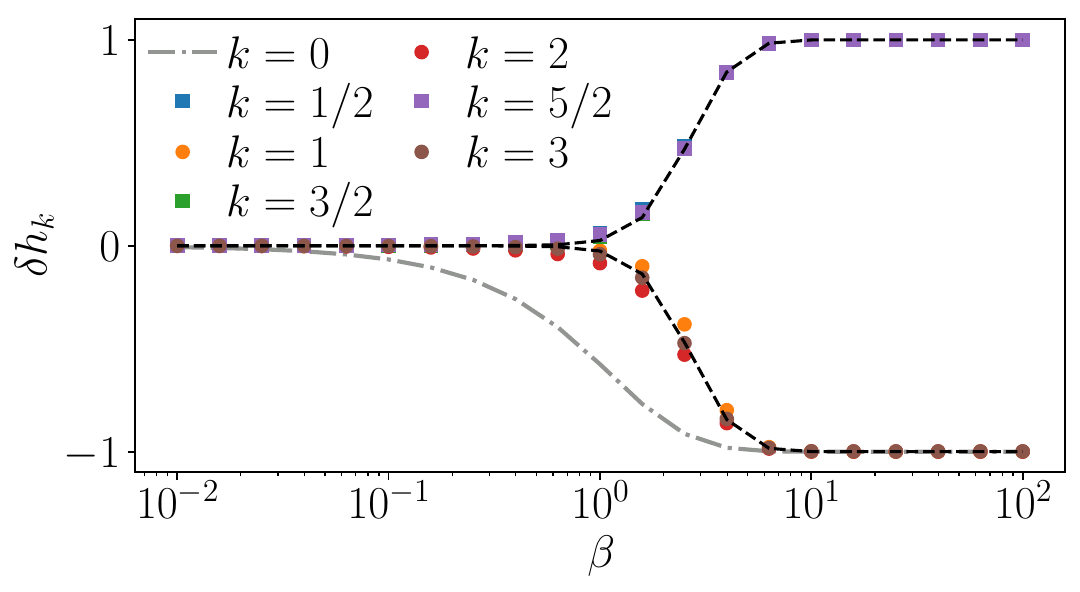}
		\caption{\small Extremal value of $\langle \hat{h} \rangle$ around $t=k\tau$ for the spin-$1$ XY model with system size $N=9$ and using $\hat{H}_\mathrm{i}$ in Eq.~\eqref{eq:Hi_XY}. The dashed black line indicates the theoretical prediction from Eq.~\eqref{eq:obs_approx}, while the dashed-dotted gray line illustrates the value of $\delta h$ at $t=0$. 
		}\label{fig:XY_obs}
	\end{figure}

	Finally, in Fig.~\ref{fig:XY_obs} we study the deviation of the expectation value of $\hat{H}_\mathrm{i}/N$ from the thermal value after the quench. Similar to the fidelity data, the numerical results match well with the theoretical prediction.
	
	Overall, we see that our approximation of non-thermalizing dynamics by only the ground state holds. In particular, the agreement with analytical prediction is good considering the relatively small system sizes accessible to the exact numerics on the spin-1 XY model. 
	Nonetheless, one might wonder about the influence of the choice of $\hat{H}_\mathrm{i}$. Indeed, choosing an operator more closely related to the algebraic structure of scars could maybe lead to different results. In particular, we can opt for a Hamiltonian such that the scarred subsapce is concentrated in the low-lying excitations. In Appendix~\ref{app:alternative}, we do this by adding a strong energy penalty on the $\ket{0}$ state to $\hat{H}_\mathrm{i}$. This lifts many states higher in energy while leaving the QMBS eigenstates untouched as they have no overlap on any configuration with $\ket{0}$ sites. As a consequence, the excited states with energy $E_n$ have $n$ $\ket{1}$ sites in a background of $\ket{-1}$, similar to the scarred eigenstates. Nonetheless, we find close results to those with $\hat{H}_\mathrm{i}$ in Eq.~(\ref{eq:Hi_XY}), meaning that only the ground state contributes to revivals even in this more optimal case.

	\section{The PXP model}\label{sec:PXP}
	
	The second model we consider is the PXP model~\cite{FendleySachdev,Lesanovsky2012} which comprises a 1D chain of spin-$1/2$ degrees of freedom,
	\begin{equation}
		\label{eq:PXPHam}
		\hat{H}_\mathrm{f,PXP}=\sum_{j=1}^N \hat{P}_{j-1}\hat{\sigma}^x_j \hat{P}_{j+1},
	\end{equation}
	defined in terms of Pauli matrices $\hat{\sigma}^\alpha$, $\alpha=x,z$, and the projector $\hat{P}_j=(\hat{\mathds{1}}_j-\hat{\sigma}_j^z)/2$. Here we assume periodic boundary conditions (PBCs), identifying $(N+1)\equiv 1$, which will allow us to reach larger system sizes and demonstrate the robustness of the results.
	
	The PXP model physically arises as an effective model of Rydberg atoms in the strong Rydberg blockade regime~\cite{Labuhn2016}. Moreover, the same model can be realized at a special resonance condition in the one-dimensional Bose-Hubbard optical lattice in the presence of tilt potential~\cite{Su2022}.  The key property of the PXP model is that neighboring excitations are forbidden due to the nearest-neighbor interactions $V_\mathrm{nn}$ (e.g., due to van der Waals forces between Rydberg atoms) being much larger than any other term in the Hamiltonian. In the spin language, this regime is implemented via the projectors $\hat{P}$ which ensure that flips do not generate any pairs of $\cdots \uparrow\uparrow \cdots$, hence the dynamics always remains within the constrained subspace. Unless specified otherwise, we will work fully within the constrained sector of the Hilbert space where there are no neighboring pairs $\cdots \uparrow\uparrow \cdots$. This implicitly assumes that the temperatures considered below should be in the regime $k_B T \ll V_\mathrm{nn}$. 
	
	The PXP model displays non-thermalizing dynamics when initialized in the N\'eel state, $\ket{\mathbb{Z}_2} \equiv \left |\uparrow\downarrow\uparrow\downarrow...\uparrow\downarrow \right \rangle$. Evolving this state with the Hamiltonian in Eq.~\eqref{eq:PXPHam}, one observes that the dynamics of local observables is approximately regular~\cite{Turner2018}. By contrast, other initial states exhibit fast equilibration, as expected in a chaotic system~\cite{Bernien2017}. 
	Conversely, this atypical dynamics is also reflected in the ergodicity breaking among a subset of eigenstates of the PXP model~\cite{Turner2018b, lin2018exact, Omiya2022}, even in the presence of perturbations~\cite{Lin2020, MondragonShem2020} or in energy transport~\cite{Ljubotina2023Transport}. 
	
	Given the special role of the $\ket{\mathbb{Z}_2}$ state for scarred dynamics in the PXP model, for our finite-temperature state preparation we use the staggered magnetization operator
	\begin{align}\label{eq:H_prep1}
		\hat{H}_\mathrm{i}=\hat{\mathds{1}}N + \hat{M}_S N, \quad \hat{M}_S = \frac{1}{N} \sum_{j=1}^N (-1)^j\hat{\sigma}^z_j,
	\end{align}
	which has the $\ket{\mathbb{Z}_2}$ state as its unique ground state with zero energy. The initial Hamiltonian~\eqref{eq:H_prep1} is chosen as it is easily realizable in experiment and quantum simulation, it breaks the degeneracy between the N\'eel state and its translated equivalent, and its first excited eigenstates can be viewed as defects on top of the $\ket{\mathbb{Z}_2}$ state due to thermal fluctuations.
	
	\begin{figure}[tb]
		\centering
		\includegraphics[width=\linewidth]{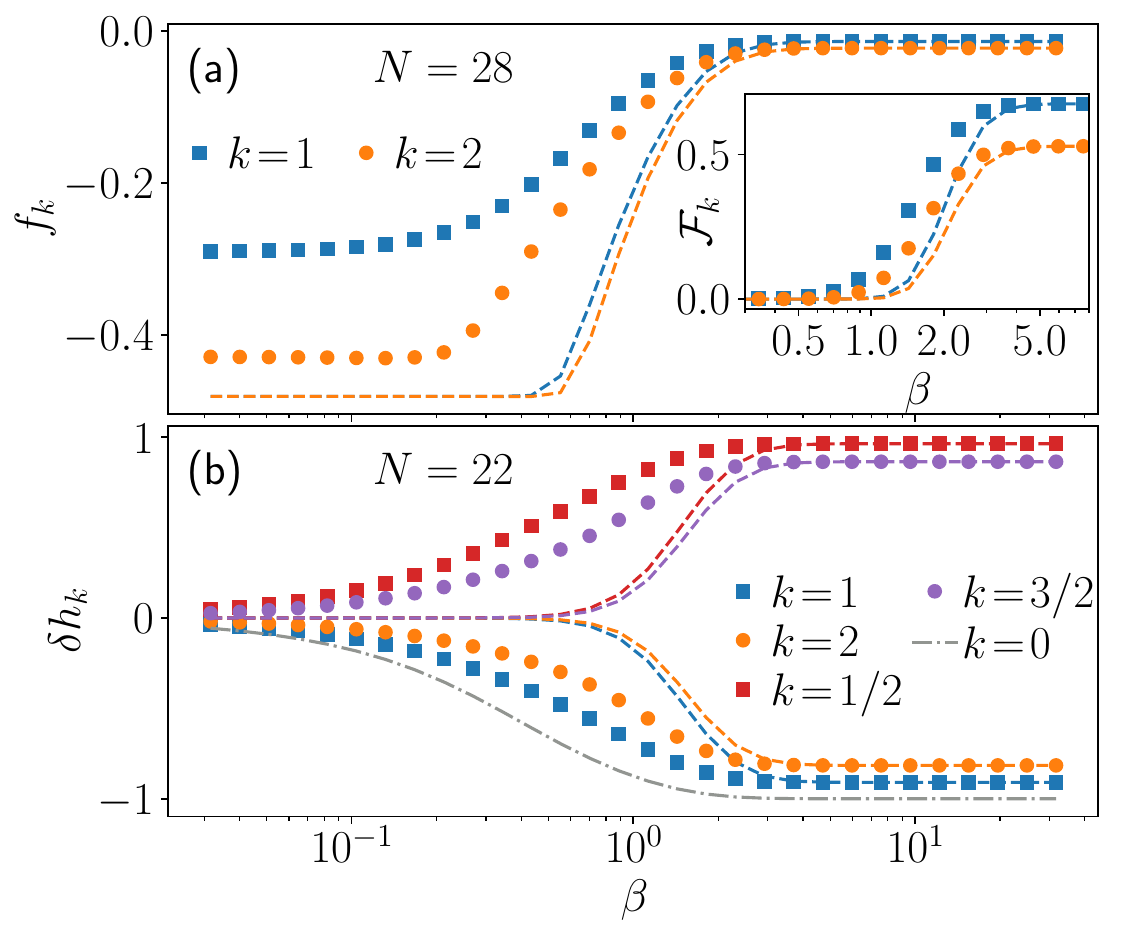}
		\caption{\small (a) Fidelity density and (b) deviation of staggered magnetization density as a function of inverse temperature in the PXP model. Inset of (a) shows the fidelity. All quantities show strong deviation from the naive expectation 
			of Eqs.~\eqref{eq:Fth} and \eqref{eq:obs_approx} denoted by the dashed lines of the same color.
			The deviation from the expected behavior should be contrasted with the spin-1 XY model in Figs.~\ref{fig:XY_fidNV} and~\ref{fig:XY_obs}. Note that the latter results for the spin-1 XY model are obtained in much smaller system sizes compared to the PXP model in this figure.
		}\label{fig:PXP_fid}
	\end{figure}
	
\begin{figure}[tb]
		\centering
		\includegraphics[width=\linewidth]{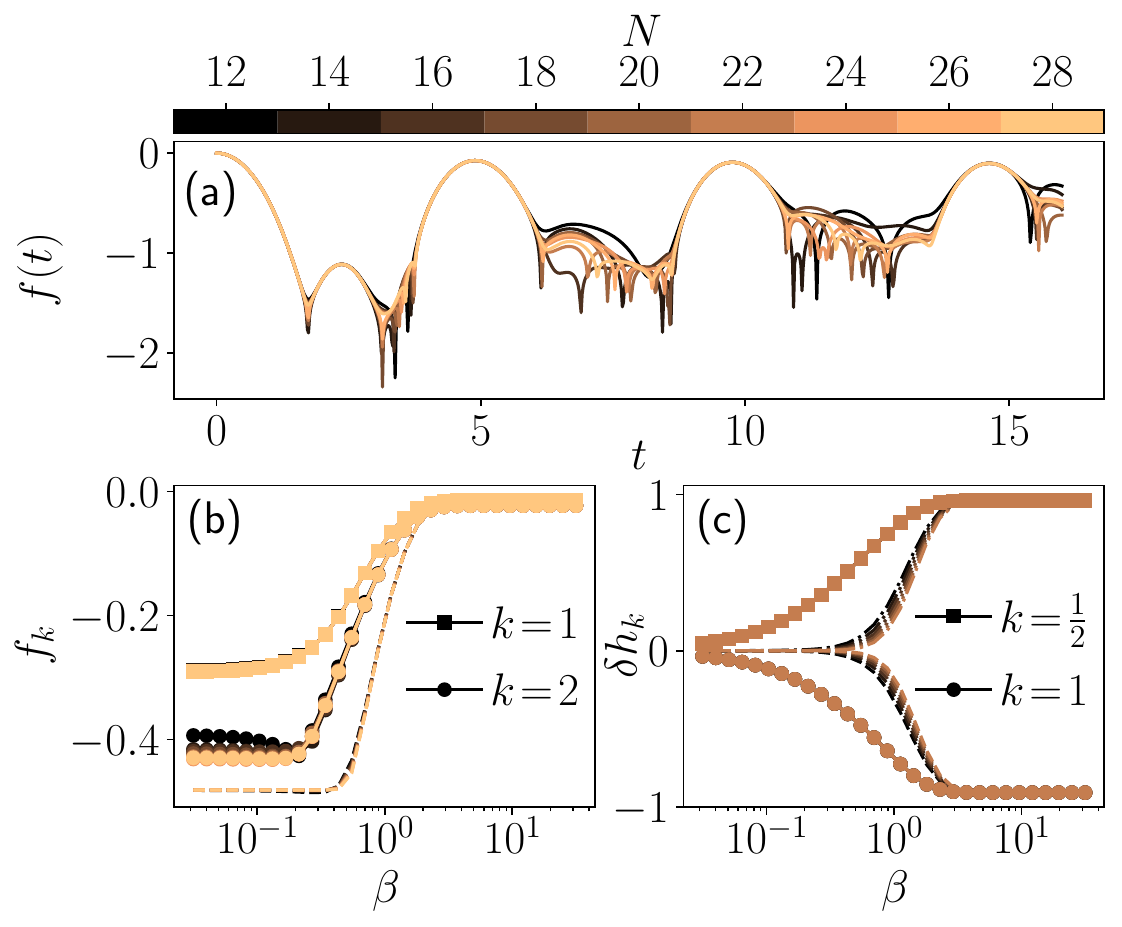}
		\caption{\small Fidelity density and observables after a finite-$T$ quench in the PXP model. (a) Fidelity density over time for different system sizes for $\beta=0.89$, away from the high- and low-temperature regimes.
			(b) Fidelity density and (c) observable extrema. The dashed lines correspond to the theoretical expectations of Eqs.~\eqref{eq:Fth} and \eqref{eq:obs_approx} for $k=1$. Both metrics are well converged in system size and show robustness to finite temperature when compared to the expected behavior.}\label{fig:PXP_scaling}
	\end{figure}

	The dynamics of $\mathcal{F}_k$, $f_k$, and $h_k$ are obtained via exact diagonalization and plotted in Fig.~\ref{fig:PXP_fid} for various system sizes indicated in the legend. For reference, we also plot the predictions of Eqs.~\eqref{eq:Fth} and~\eqref{eq:obs_approx} with dashed lines. 
	Fig.~\ref{fig:PXP_fid} shows that, for all the metrics, there are surprisingly strong deviations from theoretical predictions.
	An obvious reason for the mismatch between numerics and theoretical predictions could be finite-size effects. 
	This appears unlikely, however, as a sensitive quantity such as the fidelity density is well converged in system size, as shown in Fig.~\ref{fig:PXP_scaling} for an illustrative point $\beta \sim 1$, away from both the $\beta\approx 0$ and $\beta\to \infty$ regimes. One can clearly observe  fidelity peaks at times that are multiples of $\tau \approx4.8$, which coincides with the known revival period of the PXP model~\cite{Turner2018b}. Consequently, we still see strong deviations in $f_k$ and $\mathcal{F}_k$ at large system size $N=28$, where $\mathcal{D}=710647$, while for $h_k$ we probed system sizes up to $N=22$, where $\mathcal{D}=39603$. A detailed study of finite-size scaling of $f_k$ and $\delta h_k$ is provided in Fig.~\ref{fig:PXP_scaling}(b). These results show that both quantities are well converged already at $N\approx 20$, and we expect the observed behavior to persist in larger systems, including the larger-than-expected fidelity density near infinite temperature. Note that the corresponding results for the spin-1 XY model, Figs.~\ref{fig:XY_fidNV}-\ref{fig:XY_obs}, showed good agreement with the analytical prediction despite much smaller system sizes compared to the PXP model.
	
\begin{figure}[tb]
		\centering
		\includegraphics[width=\linewidth]{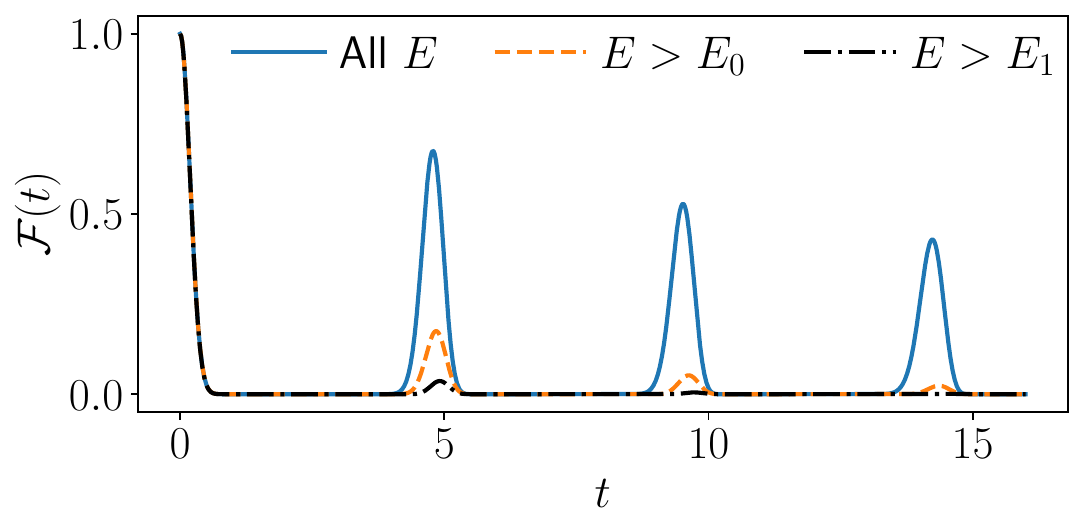}
		\caption{\small Interferometric amplitude after a quench in the pure PXP model for $N=28$ at zero temperature,  with or without energy penalties placed on the low-energy spectrum. Revivals can be seen even when the ground state (``$E{>}E_0$'') and the first set of excited states (``$E{>}E_1$'') are excluded from the initial Gibbs state.  The existence of revivals in the cases with energy penalty should be contrasted with their absence in the spin-1 XY model in Fig.~\ref{fig:XY_fid_NVGS}.
		}\label{fig:PXP_pure_no_GS}
	\end{figure}

	A more plausible explanation for the mismatch between numerics and theory in Fig.~\ref{fig:PXP_fid} would be the presence of other states leading to regular dynamics beyond the ground state of $\hat{H}_\mathrm{i}$. We test this in Fig.~\ref{fig:PXP_pure_no_GS} where we compute the fidelity in the case where the ground state is effectively projected out using Eq.~(\ref{eq:penalty}). 
	Not only are clear revivals visible when the ground state is excluded, the same is true when the first set of excitations is excluded as well. This indicates that the assumption that only the ground state yields a significant contribution is not correct in the case of PXP, accounting for the discrepancy with the theoretical predictions.
	
	This directly implies that, unlike most models with exact QMBS, the PXP model possesses additional families of scarred eigenstates with a similar structure. While this analogous structure allows them to show up in the low-energy spectrum of $\hat{H}_\mathrm{i}$, it also means that it is very difficult to remove them to try and recover the same behavior as in the XY model without destroying the main set of scarred eigenstates.
	
	\section{Quantum simulation}\label{sec:IBM}
	
	Our previous results for the PXP model strongly suggest that scarring signatures persist at finite temperature. We now demonstrate that this robustness can be witnessed in a current generation of quantum simulators. We have employed the IBM quantum processor, Kolkata, which uses a heavy hex topology and has quantum volume $128$~\cite{IBMdata}, to simulate finite-$T$ quenches in the PXP model. 
	
	The IBM processors use a cross-resonance gate to generate the CNOT entangling operation. On this hardware, we simulated the time dependence of the staggered magnetization, $\hat{M}_S$, in Eq.~\eqref{eq:H_prep1}. We simulate the evolution of the system under the Hamiltonian~\eqref{eq:PXPHam} but now, for convenience, assuming open boundary conditions. As in the classical simulations, we simulate evolution for an initial Gibbs state~\eqref{eq:rho} at temperature $1/\beta$, working fully within the constrained Hilbert space. However, rather than preparing the thermal state~\eqref{eq:rho} explicitly on the quantum computer, we use the $E\rho Oq$ method~\cite{Gustafson:2020yfe,Lamm:2018siq,Harmalkar:2020mpd,Saroni:2023uob}, which involves sampling from the density matrix~\eqref{eq:rho} via the traditional Markov Chain Monte Carlo (MCMC) method, see Appendix~\ref{app:algo} for details. We have used the suite of error mitigation techniques provided by QISKit Runtime \cite{qiskit2024}, which include: dynamic decoupling \cite{Ezzell2022,Qi:2022gdn,Morong2023, Jurcevic_2021,Niu:2022wpa,Niu:2022jnx,Mundada:2022roq,qiskit2024}, randomized compiling \cite{2016efficienttwirling,Erhard_2019,li2017efficient, 2018efficienttwirling, 2013PhRvA..88a2314G, 2016efficienttwirling,Silva-PT2008,Winick:2022scr}, and readout mitigation (specifically T-REx) \cite{PhysRevApplied.14.054059,PhysRevApplied.12.054023,PhysRevApplied.10.034040,Sarovar2020detectingcrosstalk,Berg:2020ibi,Smith:2021iwt,Rudinger:2021nhd,10.1145/3352460.3358265,Harrigan2021,PhysRevA.100.052315,Maciejewski2020mitigationofreadout,Nachman2020,Hicks:2021uvm,PhysRevLett.122.110501,PhysRevA.101.032343,Hamilton:2020xpx,Geller_2021,PhysRevLett.119.180511}. 
	We also used a rescaling procedure to counteract the signal loss from the effective depolarizing channel caused by the randomized compiling \cite{Urbanek2021,Vovrosh:2021ocf,ARahman:2022tkr}.
	
	\begin{figure}[tb]
		\includegraphics[width=\linewidth]{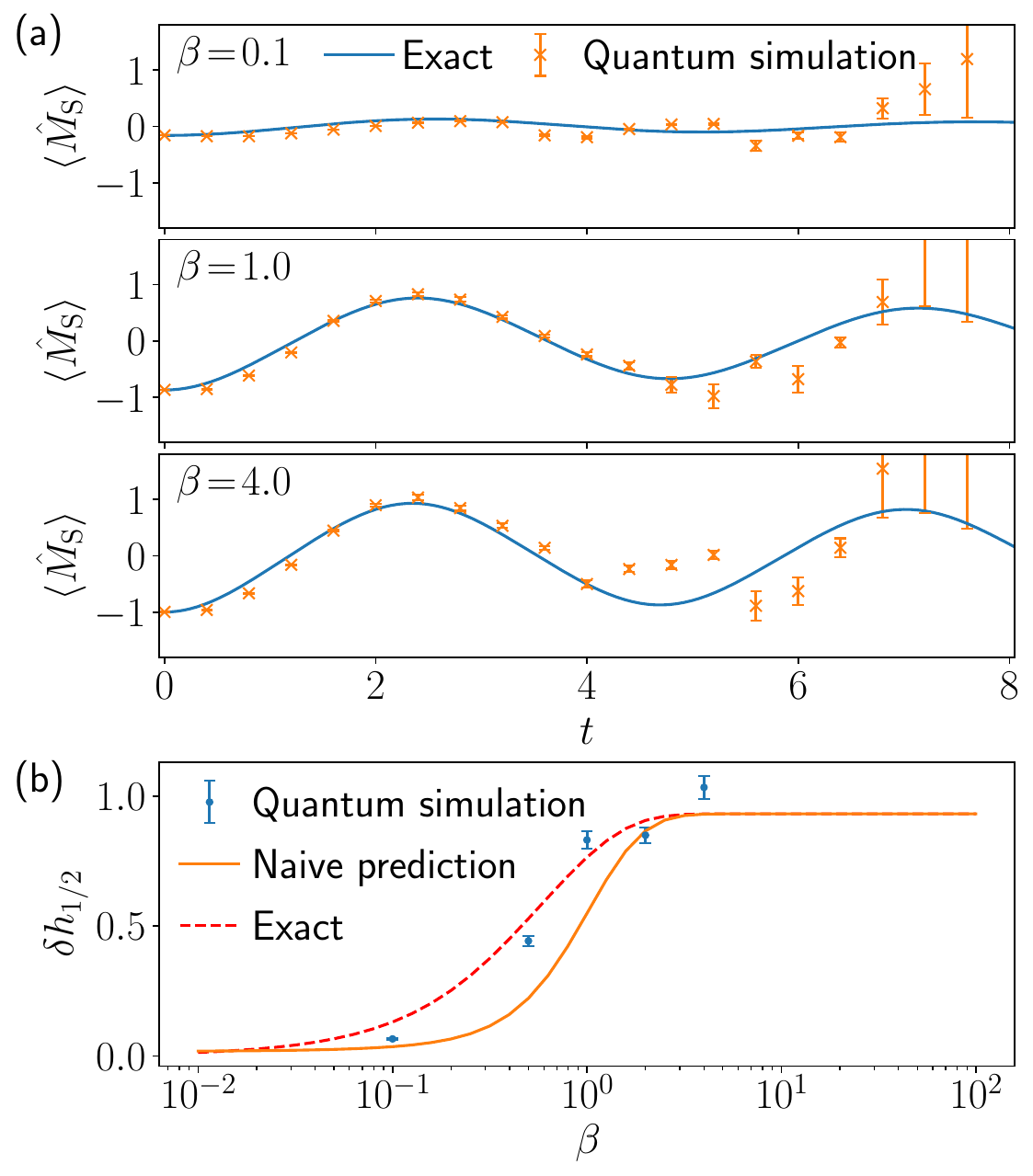}
		\caption{\small IBM quantum simulation of finite-$T$ quenches in the PXP model. (a) Time dependence of $\hat{M}_\mathrm{S}$ for $\beta\in\lbrace 0.1, 1, 4\rbrace$ in the PXP model with $8$ qubits and OBC. The data obtained on the IBM device shows good agreement with the numerical simulations. 
			(b) Relative deviation after a half period of $\langle \hat{H}_\mathrm{i} \rangle/N$ from its predicted value of Eq.~\eqref{eq:obs_approx}. The quantum simulation data shows larger deviation than the naive expectation. In both panels, the error bars correspond to partial systematic errors from rescaling and statistical sampling from the density matrix. The simulations for $\beta=2$ were run on July 26, 2023 and all other $\beta$ on August 4th. }
		\label{fig:simul_dh}
	\end{figure}
	
	We have simulated the PXP model at $5$ different inverse temperatures $\beta\in\lbrace 0.1, 0.5, 1, 2, 4\rbrace$ and generated $N_c = 10^5$ configurations at each $\beta$. The time-evolution operator was decomposed using the Trotter approximation with a time step of $\delta t = 0.4$.   
	The time evolution of $\hat{M}_\mathrm{S}$ is shown for $8$ qubits in Fig.~\ref{fig:simul_dh}(a) for $\beta\in\lbrace 0.1, 1, 4\rbrace$. While this is a relatively small system, the PXP model is known to be difficult to simulate even with advanced error-mitigation techniques~\cite{Chen2022Error}.
	We find that a reliable signal for the time dependence can be obtained up to one oscillation or roughly ten Trotter steps. 
	From this data, we can extract $\delta h_{k}$; in fact, as $\langle \hat{h} \rangle_{\beta=0}=1$, it is straightforward to see that $\delta \hat{h}=\hat{M}_\mathrm{S}$. In Fig.~\ref{fig:simul_dh}(b), we plot $\delta h_{1/2}$ for our quantum simulation along with the exact results. Both show larger deviations from the thermal value that our naive expectation would predict in Eqs.~\eqref{eq:Fth} and \eqref{eq:obs_approx}.  
	While we need to keep in mind the small system size used, which limits the accuracy of our prediction based solely on the ground-state contributing, its relatively good agreement with the exact data means that we can expect the same kind of behavior in larger systems.
	
	\section{Conclusions and discussion}\label{sec:conc}
	
	We have studied the fate of wave function revivals at finite temperature in two classes of QMBS models. The initial density matrix at temperature $1/\beta$ is produced by an annealing procedure, instead of being a pure state previously considered in the literature.
	For the spin-1 XY model, a prototype model with exact QMBS states, we find that the low-$T$ behavior of the return fidelity as well as the oscillations of observables are  well described by the analytical toy model, Eqs.~\eqref{eq:Fth}-\eqref{eq:obs_approx}, implying that only the $T=0$  state gives a non-vanishing contribution and the QMBSs are fragile with respect to finite temperature.   
	On the other hand, we found that the PXP model violates this expectation and displays unexpectedly robust signatures of QMBS dynamics even at high temperatures.  In contrast to previous work that focused on the efficient preparation of QMBS eigenstates~\cite{Gustafson:2023lhl},
	we have observed robust QMBS signatures at finite temperature in the dynamics of both the fidelity and local observables in the PXP model. Finite-size scaling shows that this behavior is well converged within the accessible system sizes.  Using a digital quantum computer, we have demonstrated persistent QMBS revivals in the IBM device at finite temperature. 
	
	The main question stemming from our study is how to understand the striking difference in the finite-$T$ behavior of two scarred models, the PXP and spin-1 XY model. In an attempt to reconcile their different behaviors, in Appendix~\ref{app:PXPpert} we studied finite-$T$ quenches for a perturbed variant of the PXP model which is known to exhibit nearly perfect scarring~\cite{Khemani2018,Choi2018,Bull2020}. Due to the nearly-perfect QMBS subspace, the perturbed PXP model may be expected to behave similarly to the spin-1 XY model. Nevertheless, our results suggest that this expectation is incorrect, as the finite-$T$ dynamics in the perturbed PXP model is found to be similar to the \emph{unperturbed} PXP results presented above.
	
	We attribute the difference in finite-$T$ behavior between the PXP and other models to the different algebraic properties of their QMBS states. Namely, the QMBS states in the PXP model form a representation of a large $\mathrm{su}(2)$ spin~\cite{Choi2018}, which is a special case of the  ``restricted spectrum generating algebra'' that describes many other QMBS models, including the spin-$1$ XY magnet~\cite{MoudgalyaReview,Buca2019,MarkLinMotrunich,Dea2020,Buca2023}. In most of these models, the non-thermal eigenstates are completely decoupled from the thermal bulk, hence they \emph{exactly} form a \emph{single} algebra representation. By contrast, in the PXP model the algebra is inexact, due to the small residual couplings to the thermal bulk. More importantly, the non-thermal eigenstates form towers of \emph{multiple} $\mathrm{su}(2)$ representations that originate from a collective spin-$1$ degree of freedom~\cite{Omiya2022}. This means that when starting from a finite-temperature ensemble in the PXP model, we can have coherent contributions from states belonging to different $\mathrm{su}(2)$ representations, which effectively provides a stronger ``shield'' for QMBSs against finite temperature. 
	
	Unfortunately, due to a lack of an exhaustive construction of multiple $\mathrm{su}(2)$ representations in the PXP model, their impact on finite-$T$ quench dynamics remains a conjecture at this stage. One interesting direction to pursue would be to construct toy models with a controllable number of embedded algebra representations and probe their finite-$T$  behavior. On the other hand, it is worth noting that there are also other frameworks for building QMBS models that extend beyond the simple Lie algebra scheme considered here, e.g., ~\cite{Dea2020, Pakrouski2020, Ren2021, Moudgalya2022}, and it would be interesting to understand if any of them display a similar robustness to finite temperature.
	
	\begin{acknowledgments}
		J.-Y.D. and Z.P. acknowledge support by 
		the Leverhulme Trust Research Leadership Award RL-2019-015 and  EPSRC grants EP/R513258/1, EP/W026848/1. 
		Statement of compliance with EPSRC policy framework on research data: This publication is theoretical work that does not require supporting research data. This research was supported in part by grant NSF PHY-2309135 to the Kavli Institute for Theoretical Physics (KITP). J.C.H.~acknowledges funding from the European Research Council (ERC) under the European Union’s Horizon 2020 research and innovation programm (Grant Agreement no 948141) — ERC Starting Grant SimUcQuam, and by the Deutsche Forschungsgemeinschaft (DFG, German Research Foundation) under Germany's Excellence Strategy -- EXC-2111 -- 390814868. 
		This material is based upon work supported by the U.S. Department of Energy, Office of Science, National Quantum Information Science Research Centers, Superconducting Quantum Materials and Systems Center (SQMS) under the contract No. DE-AC02-07CH11359.  E.J.G.~was supported by the NASA Academic Mission Services, Contract No. NNA16BD14C. 
		This research used resources of the Oak Ridge Leadership Computing Facility, which is a DOE Office of Science User Facility supported under Contract DE-AC05-00OR22725.
		We acknowledge the use of IBM Quantum services for this work. The views expressed are those of the authors, and do not reflect the official policy or position of IBM or the IBM Quantum team.
	\end{acknowledgments}

	\appendix

	\section{Effective temperature with respect to the quench Hamiltonian $\hat{H}_\mathrm{f}$}\label{app:temp_fin}
	
	As our focus is on QMBSs, we want to make sure that there are no other sources of periodic dynamics. In particular, we want to avoid the initial state being close to the ground (or ceiling) state of $\hat{H}_\mathrm{f}$, as approximately periodic dynamics could arise from the initial state having large overlap with only the ground state and a few low-lying excited states. Thus, we make sure that the initial density matrix $\hat{\rho}(\beta)$ that we consider is always essentially at infinite temperature with respect to $\hat{H}_\mathrm{f}$, which can be expressed as a condition $\Tr\left[\hat{\rho}(\beta)\hat{H}_\mathrm{f}\right]\approx \Tr[\hat{H}_\mathrm{f}]/\mathcal{D}$ for all $\beta$, with $\mathcal{D}$ denoting the Hilbert space dimension. We emphasize that this statement is only about the energy expectation value of $\hat{H}_\mathrm{f}$ with respect to $\hat{\rho}$ matching that of infinite temperature, and not necessarily about $\hat{\rho}$ being a Gibbs state for $\hat{H}_\mathrm{f}$.  
	
	\subsection{PXP model}
	
	The spectrum of $\hat{H}_\mathrm{f,PXP}$ is symmetric around $E=0$ as this Hamiltonian anticommutes with $\hat{K}_\mathrm{PXP}=\prod_{j=1}^N \hat{\sigma}^z_j$. The energy eigenvalues must then sum to zero since for any energy eigenstates $\ket{E_{f,k}}$ of  $\hat{H}_\mathrm{f,PXP}$ with energy $E_{f,k}\neq 0$ there exists another eigenstate $\ket{E_{f,j}}=\hat{K}_\mathrm{PXP}\ket{E_{f,k}}$ which has an energy $E_{f,j}=-E_{f,k}$. This implies that, at infinite temperature, the effective energy is 0, since it is equal to the mean of all energy eigenvalues. At the same time, the initial state considered is $\ket{\psi_0}=\left |\uparrow\downarrow\uparrow\downarrow...\uparrow\downarrow \right \rangle$. Importantly, this state is a basis state in the computational basis while the PXP Hamiltonian $\hat{H}_\mathrm{f,PXP}$ is purely off-diagonal in this basis. This means that $\langle \psi_0|\hat{H}_\mathrm{f,PXP}|\psi_0\rangle=0$, which corresponds to  infinite temperature.
	
	In fact, the previous argument  can be applied to all states $\hat{\rho}$ considered in this work to show they effectively correspond to infinite temperature. Indeed, since $\hat{H}_\mathrm{i,PXP}$ is diagonal in the computational basis, we can take its basis states as eigenstates $\ket{E_{i,k}}$. Then, for \emph{all} these eigenstates, we must have $\langle E_{i,k}|\hat{H}_\mathrm{f,PXP}|E_{i,k}\rangle=0$, for the same reason as for the $\ket{\psi_0}$ state above. This directly implies that 
	\begin{equation}
		\Tr\left[\hat{\rho}(\beta)\hat{H}_\mathrm{f,PXP}\right]{=}\frac{1}{Z}\sum_k e^{-\beta E_{i,k}}\langle E_{i,k}|\hat{H}_\mathrm{f,PXP}|E_{i,k}\rangle{=}0.
	\end{equation}

	Finally, we note that the perturbation considered in Appendix \ref{app:PXPpert} is also off-diagonal in the computational basis and preserves the anti-symmetry of the spectrum. Thus, the same conclusions also apply to the perturbed PXP model.

	\subsection{Spin-1 XY model}
	
	In the case with $D=0$, $\hat{H}_{f,XY}$ anticommutes with the operator $\hat{K}_\mathrm{XY}=\hat{\sigma}^x_1\hat{\sigma}^y_2\hat{\sigma}^x_3\hat{\sigma}^y_4\cdots \hat{\sigma}^x_{N-1}\hat{\sigma}^y_N$. As in the PXP case, this implies that the spectrum is symmetric around 0 and that the energy expectation value at infinite temperature is $E=0$.

	The expectation value of the state $\ket{\psi_0}$ under $\hat{H}_\mathrm{f,XY}$ can be computed analytically and is given by $DN$. So for $D=0$ we have that its energy is 0 and so that this state is at infinite temperature. We have a similar result for $\hat{\rho}$ at all values of $\beta$. Indeed, the eigenvalues of $\hat{H}_\mathrm{i,XY}$ can all be written as basis states in the basis $\{\ket{-},\ket{0},\ket{+} \}$, with 
	$\ket{+}=\left(\ket{+1}+\ket{-1}\right)/\sqrt{2}$ and $\ket{-}=\left(\ket{+1}-\ket{-1}\right)/\sqrt{2}$. 
	This directly implies that they all have zero net magnetization as this is the case for $\ket{-}$, $\ket{0}$ and $\ket{+}$. Thus, their energy contribution from the term $h\sum_{j=1}^N \hat{S}^z_j$ is 0. It is also straightforward to check  that all the other terms (for $D=0$) are purely off-diagonal in the basis $\{\ket{-},\ket{0},\ket{+} \}$ and so that all eigenstates of $\hat{H}_\mathrm{i,XY}$ have expectation value $E=0$ under $ \hat{H}_\mathrm{f,XY}$. As in the PXP case, this directly implies that $\Tr\left[\hat{\rho}(\beta)\hat{H}_\mathrm{f,XY}\right]=0$ for all $\beta$.

	In order to remove as much structure as possible and match the discussion of QMBSs in the spin-1 XY model in Ref.~\cite{Iadecola2019_2}, we now set $D=0.1$. As two other parameters ($h$ and $J$) are equal to 1,  the value $D=0.1$ should not cause significant shifts in the spectrum while being large enough to break the anti-symmetry. In fact, one can analytically show that the infinite temperature energy is shifted to $2DN/3$. It is important to remember that for such a local 1D system, the range of the spectrum also scales linearly with system size (up to a subleading correction), thus the relevant quantity is the energy density, $E/N$. As the system size is varied, we see that the difference in infinite-temperature energy density between $D=0$ and $D=0.1$ is constant, $2D/3=1/15$, which is relatively small compared to the full energy-density range. We numerically compute its minimum and maximum values, which we find to be $-2.3513$ and $2.3513$ for $D=0$, and $-2.3012$ and $2.4019$ for $D=0.1$ (for system size $N=14$, with similar differences in smaller system sizes). This illustrates that the effect on the spectrum by changing $D=0 \to 0.1$ is small.
	
	A similar analysis can be performed for the state $\ket{\psi_0}$, for which the energy density is $D$. The difference between this energy density and that of infinite temperature is constant at $D/3=1/30$. Once again, this difference is small when compared to the full range of the energy density ($-2.3012$ to $2.4019$ for $D=0.1$ at $N=14$).
	
\begin{figure}[t!]
		\centering
		\includegraphics[width=\linewidth]{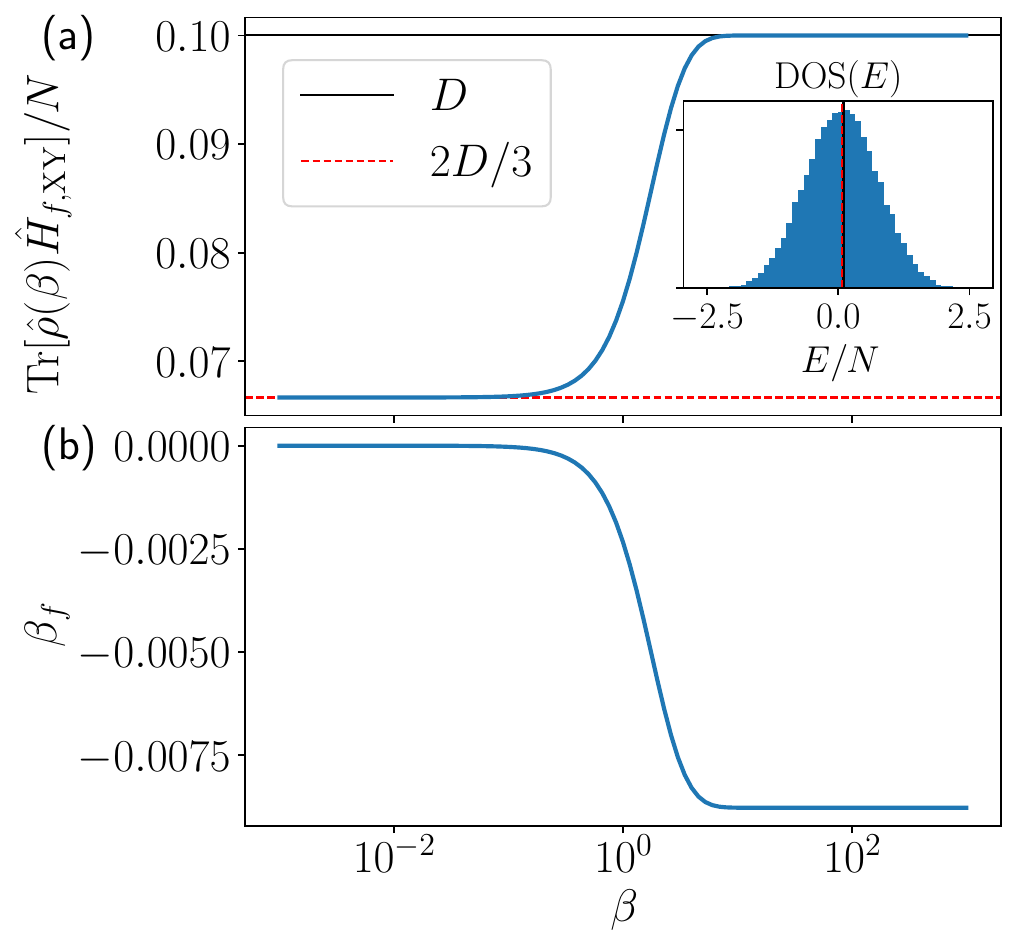}
		\caption{
			\small Position of the energy of $\hat{\rho}(\beta)$ in the spectrum of $\hat{H}_\mathrm{f,XY}$ as $\beta$ is varied in the spin-1 XY model for $N=8$. (a) Energy density expectation of $\hat{\rho}(\beta)$ with respect to $\hat{H}_\mathrm{f,XY}$. The solid black line and the red dashed line represent the energy density at $\beta=\infty$ and $\beta=0$, respectively, in both the main plot and the inset. The inset shows the density of states (DOS) of $\hat{H}_\mathrm{f,XY}$. The fact that the two lines are almost superimposed when looking at the full energy density range shows that $\hat{\rho}$ is essentially in the middle of the spectrum for all $\beta$. (b) Effective temperature $\beta_f$ of $\hat{\rho}(\beta$) with respect to $\hat{H}_\mathrm{f,XY}$.
		}
		\label{fig:XY_betaf}
	\end{figure}

	In Fig.~\ref{fig:XY_betaf}(a), we plot the energy density of $\hat{\rho}$ with respect to $\hat{H}_\mathrm{f,XY}$ for the full range of $\beta$ that we investigate. As expected, the energy density varies  smoothly between $2D/3$ and $D$. The inset illustrates that this range is small when compared to the full spectrum by displaying $D$ and $2D/3$ over the density of states (DOS).

	The effective inverse temperature $\beta_f$ of $\hat{\rho}$ with respect to $\hat{H}_\mathrm{f,XY}$ is defined by the condition
	\begin{equation}
		\Tr\left[\hat{\rho}\hat{H}_\mathrm{f,XY}\right]=\frac{\Tr\left[\hat{H}_\mathrm{f,XY}e^{-\beta_f \hat{H}_\mathrm{f,XY}}\right]}{\Tr\left[e^{-\beta_f \hat{H}_\mathrm{f,XY}}\right]}.
	\end{equation}
	As $\hat{\rho}$ depends on $\beta$, the same is true for $\beta_f$. However, as expected, $\beta_f$ is always small. For $N=8$, we find that it varies between 0 (for $\beta=0$) and 
	$-8.8 \times 10^{-3}$ (for $\beta=\infty$). This is illustrated in Fig.~\ref{fig:XY_betaf}(b). We emphasize that this value of $\beta_f$ is even smaller than the smallest values of $\beta$ that we consider for state preparation, for which no changes are seen when $\beta$ is further decreased.

	In conclusion, we find that for $D=0.1$ the density matrix $\hat{\rho}$ is effectively at infinite temperature for all values of $\beta$. As such, we expect it to thermalise quickly, unless there is an obstruction to this due to QMBSs. We briefly note that the same considerations apply when using the alternative preparation Hamiltonian discussed in Appendix~\ref{app:alternative}. Indeed, the Hamiltonian in Eq.~\eqref{eq:Hi_XY2} shares the same eigenstates as the one in Eq.~\eqref{eq:Hi_XY}, only their energy is modified. As such, the energy density of $\hat{\rho}$ still varies between $2D/3$ and $D$, leading to the same $\beta_f$. While the exact form of $\beta_f(\beta)$ might be slightly different, it will still be extremely close to infinite temperature.
	
	\section{Low-temperature approximation}\label{app:LowT}
	
	Here we derive the expected behavior of the interferometric Loschmidt echo and of the expectation value of a local observable $\hat{h}$ in a scarred system following a quench.
	
	\subsection{Interferometric Loschmidt echo}\label{app:fid}
	
\begin{figure}[t!]
		\centering
		\includegraphics[width=\linewidth]{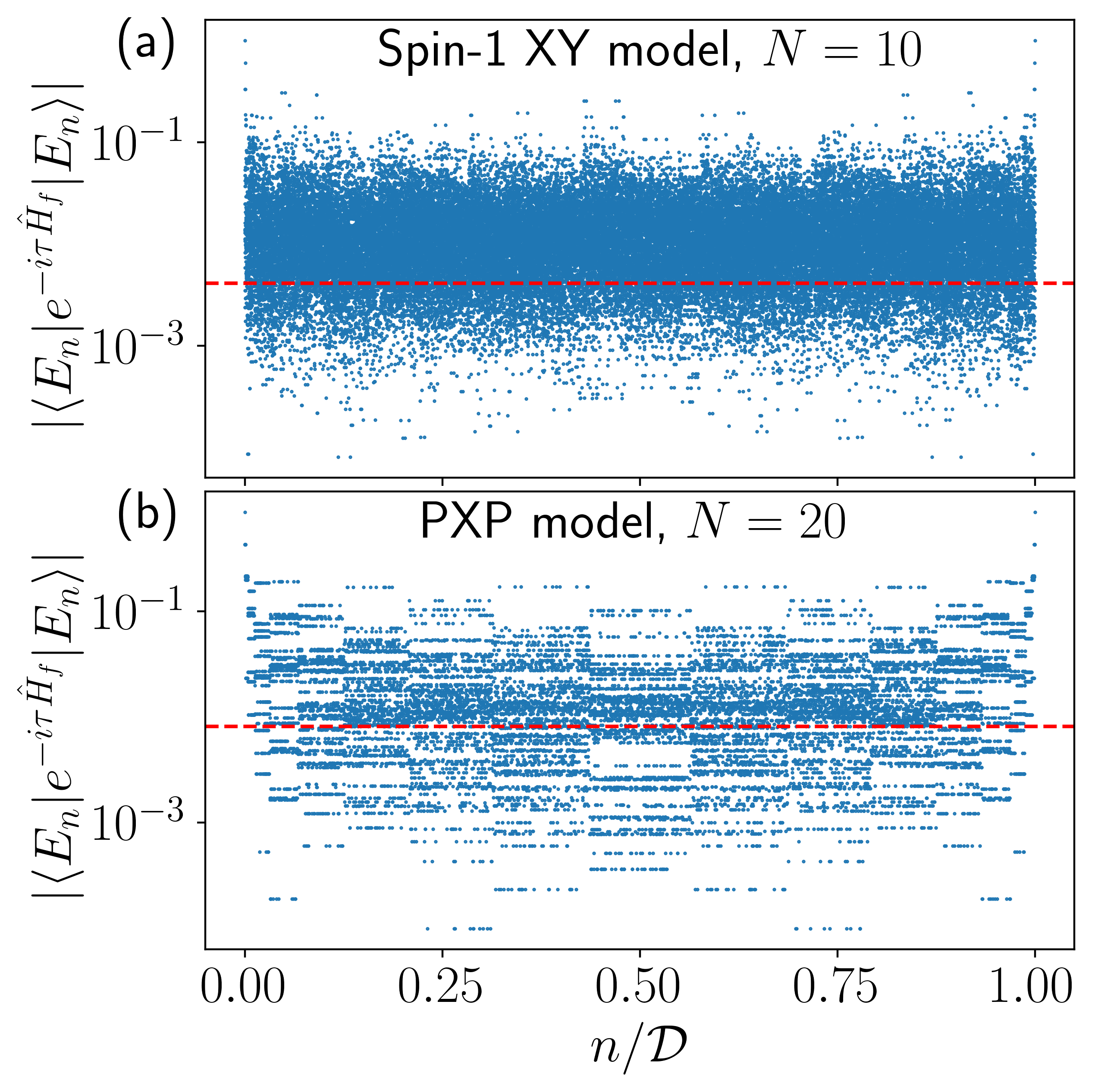}
		\caption{\small 
			Square root of the return fidelity after evolving the eigenstates $\ket{E_n}$ of $\hat{H}_\mathrm{i}$ with  $\hat{H}_\mathrm{f}$ for a single period. The eigenstates are sorted by increasing energy such that $E_{n}\leq  E_{n+1}$. The red dashed lines indicate $1/\sqrt{\mathcal{D}}$. Results shown are for (a) the spin-1 XY model with $N=10$ spins and (b) the PXP model with $N=20$. The multiple points at the same value in the latter are due to $\hat{H}_\mathrm{f,PXP}$ being invariant under translation because we use periodic boundary conditions. Overall, we see the expected behavior that only states with $n/\mathcal{D}$ close to 0 and 1 (i.e. near the ground and ceiling states) have a return fidelity close to 1. Between these two extremes, the fidelity is lower and gets even lower at longer times. However, we also see that $1/\sqrt{D}$ is a relatively crude estimate of the actual fidelity values for these states.
		}
		\label{fig:eig_fid}
	\end{figure}

	Let us first focus on the interferometric Loschmidt echo, defined in Eq.~(\ref{eq:LE}) in the main text. Let us denote by $\ket{E_n}$ the eigenstates of $\hat{H}_\mathrm{i}$ with eigenenergies $E_n$.
	As done in the main text, we will assume $E_0=0$. For a given value of the inverse temperature $\beta$ (with respect to $\hat{H}_\mathrm{i}$), our initial mixed state will then be given by 
	\begin{align}
		\hat{\rho}=\frac{1}{Z}\sum_{E_n}e^{-\beta E_n}\ketbra{E_n},
	\end{align}
	where $	Z=\sum_{E_n}\exp(-\beta E_n)$ is the partition function.
	Substituting the expression for the density matrix, Eq.~\eqref{eq:LE} becomes 
	\begin{align}
		\mathcal{F}_{\hat{\rho}}(t)=\frac{1}{Z^2}\bigg\lvert\sum_{E_n} e^{-\beta E_n}\langle E_n|e^{-i\hat{H}_\mathrm{f}t}|E_n\rangle\bigg\rvert^2.
	\end{align}
	At times that are multiples of the period, $t=k\tau$, we know that $\langle E_0|e^{-i\hat{H}_\mathrm{f}t}|E_0\rangle= \sqrt{\mathcal{F}^\infty_k}$, where the $0$ subscript denotes infinite $\beta$ (or equivalently zero temperature). 
	Let us discuss the other eigenstates $\ket{E_n}$ with $n\neq 0$. We assume they are thermalizing with respect to $\hat{H}_\mathrm{f}$, and so we should get
	\begin{eqnarray}
		\langle E_n|e^{-i\hat{H}_\mathrm{f}t}|E_n\rangle\approx e^{-ir_n}/\sqrt{\mathcal{D}},
	\end{eqnarray}
	for sufficiently long $t$. 
	This is a relatively crude approximation after a single period, however the contrast between the ground state and the rest of the spectrum is already clear at that point, as shown in Fig.~\ref{fig:eig_fid}.
	
	As the factors $e^{-ir_n}$ are essentially  random phases with a very small individual contribution, we can forget about their weights and consider an equal superposition. In practice, we assume the same prefactor for all states except the ground state. This prefactor is computed by taking the total sum of the $e^{-\beta E_n}$ for $n\neq 0$ (i.e. excluding the ground state), which is simply $Z-e^{-\beta E_0}=Z-1$ since we have set $E_0=0$. We then divide by the number of states $\mathcal{D}-1 \approx \mathcal{D}$ in the limit of large system sizes (i.e., we can neglect the exclusion of the ground state in the total count).  We can verify that this expression indeed gives 0 at $\beta=\infty$ (where only the term of the ground state should contribute) and 1 in the thermodynamic limit for $\beta=0$ (where the contribution of the ground state is irrelevant). While this approximation is not very accurate for larger values of $\beta$ where the weights of the different $E_n$ can strongly vary, in that regime the contribution of the ground state completely dominates around $t=k\tau$. As such, any inaccuracy in the contribution of the other eigenstates will be effectively negligible. On the other hand, for small $\beta$ the ground state no longer dominates but the prefactor of each eigenstate is close to equal. Thus, our approximation is justified and we can rewrite 
	\begin{equation}
		\begin{aligned}
			\mathcal{F}_{\hat{\rho}}(k\tau)&=\frac{1}{Z^2}\bigg\lvert\sqrt{\mathcal{F}^\infty_k}+\frac{Z-1}{\mathcal{D}}\sum_{n\neq 0}\langle E_n|e^{-iH_fk\tau}|E_n\rangle\bigg\rvert^2 \\
			&\approx \frac{1}{Z^2}\bigg\lvert\sqrt{\mathcal{F}^\infty_k}+\frac{Z-1}{\mathcal{D}}{\rm Tr}\left[e^{-iH_fk\tau}\right]\bigg\rvert^2.
		\end{aligned}
	\end{equation}
	Taking the expectation value, the cross product vanishes as its expectation value is zero for a chaotic system. Meanwhile $\frac{1}{\mathcal{D}^2}|{\rm Tr}\left[e^{-iH_fk\tau}\right]|^2$ is simply the spectral form factor (SFF). At times on the order of the Heisenberg time, this quantity is known to saturate to
	\begin{equation}
		\frac{1}{\mathcal{D}^2}\mathbf{E}\left[|{\rm Tr}\left[e^{-iH_fk\tau}\right]|^2 \right] \xrightarrow{k\gg 1} 1/\mathcal{D}.
	\end{equation}
	The symbol $\mathbf{E}$ denotes an expectation value (in the probabilistic sense) over several realizations with random disorder in some of the parameters. This is because the SFF is not self-averaging in general~\cite{Prange1997SFF}. However, for the PXP model considered in the main there is no known parameter that can be modified without affecting the scarring. As a consequence, we take the approximation that a single realization will be close to the expectation value, leaving us with
	\begin{equation}\label{eq:fid_k}
		\mathcal{F}_{\hat{\rho}}(k\tau)=\frac{\mathcal{F}^\infty_k}{Z^2}
		+\frac{\left(1-\frac{1}{Z}\right)^2}{\mathcal{D}}.
	\end{equation}
	for a long enough $k\tau$.
	While this approximation is relatively crude, it still provides us with a prediction for the overall magnitude of $\mathcal{F}_k$ for $\beta<1$, where the second term is the main contribution. In the regime $\beta>1$, the contribution $\frac{\mathcal{F}^\infty_k}{Z^2}$ from the ground state will instead be the only relevant one. Thus we expect a much higher accuracy in that regime and we focus on it.
	
	\subsection{Observables}\label{app:obs}
	
	We now derive an expression for the expectation value of $\hat{h}$ if only the ground state of $\hat{H}_\mathrm{i}$ shows perfect revival in $\hat{H}_\mathrm{f}$ and all other eigenstates thermalize rapidly.  As in the previous section, we denote by $\ket{E_n}$ the eigenstates of the pre-quench Hamiltonian $\hat{H}_\mathrm{i}$ and by $\tau$ the revival period. The previous assumption then translates into the statement
	\begin{equation}\label{eq:assum_obs}
		\braket{E_n|e^{-i\hat{H}_\mathrm{f} \tau}|E_m}=\delta_{n,0}\delta_{m,0}+\frac{(1-\delta_{n,0})(1-\delta_{m,0})}{\sqrt{\mathcal{D}-1}},
	\end{equation}
	with $\mathcal{D}$ the Hilbert space dimension and $\tau$ is assumed to be large. We have checked the validity of Eq.~(\ref{eq:assum_obs}) numerically finding that the magnitudes of matrix elements are indeed distributed around $1/\sqrt{\mathcal{D}}$, albeit with a large spreading, with the dominant matrix element coming from the ground state ($n=m=0$). 
	
	Eq.~(\ref{eq:assum_obs}) will prove useful to compute the expectation value of $\hat{h}=\frac{1}{N}\hat{H}_\mathrm{i}$ over time, defined as 
	\begin{equation}
		\begin{aligned}
			&\braket{\hat{h}}(t)=\Tr\left[{\hat{\rho}(t)\hat{h}}\right]\\
			&=\frac{1}{N}\sum_{n,m} P(E_m) \bra{E_n}e^{-i\hat{H}_\mathrm{f}t}\ket{E_m}\bra{E_m}e^{i\hat{H}_\mathrm{f}t}\hat{H}_\mathrm{i}\ket{E_n} \\
			&=\sum_{n,m} P(E_m) \tilde{E}_n \big\lvert\bra{E_n}e^{-i\hat{H}_\mathrm{f}t}\ket{E_m}\big\rvert^2,
		\end{aligned}
	\end{equation}
	with $P(E_n)=e^{-\beta E_n}/Z$ and $\tilde{E}_n=E_n/N$. For $t=k\tau$ we can use the assumption made in Eq.~\eqref{eq:assum_obs} to get
	\begin{equation}
		\begin{aligned}
			\braket{\hat{h}}(k\tau)&\approx P(E_0)\tilde{E}_0+\sum_{n\neq 0}\sum_{m\neq 0}\frac{P(E_n)\tilde{E}_m}{\mathcal{D}-1} \\
			&=P(E_0)\tilde{E}_0{+}\left(\sum_{m\neq 0}P(E_m)\right)\left(\frac{\sum_{n\neq 0}\tilde{E}_n}{\mathcal{D}-1}\right).
		\end{aligned}
	\end{equation}
	Using $\sum_{m\neq 0}P(E_m) {=} 1-P(E_0)$ and $\sum_n \tilde E_n = \mathcal{D}\braket{\hat{h}}_{\beta=0}$, we finally arrive at
	\begin{equation}
		\begin{aligned}
			\braket{\hat{h}}(k\tau) &\approx 
			P(E_0)\tilde{E}_0{+}\left(1{-}P(E_0)\right)\left(\frac{\mathcal{D}\braket{\hat{h}}_{\beta=0}{-}\tilde{E}_0}{\mathcal{D}{-}1}\right). \\
		\end{aligned}
	\end{equation}
	In the limit of large system sizes, we can take $\frac{\mathcal{D}}{\mathcal{D}-1}\to 1$ and $\frac{\tilde{E}_0}{\mathcal{D}-1}\to 0$, since $\tilde{E}_0=E_0/N$ is $\mathcal{O}(1)$. This leads to 
	\begin{equation}
		\begin{aligned}
			\braket{\hat{h}}(k\tau)&=P(E_0)\tilde{E}_0+\left(1-P(E_0)\right)\braket{\hat{h}}_{\beta=0}\\
			&=\braket{\hat{h}}_{\beta=0}+P(E_0)\left(\tilde{E}_0-\braket{\hat{h}}_{\beta=0}\right)\\
			&=\braket{\hat{h}}_{\beta=0}+\frac{1}{Z}\left(\tilde{E}_0-\braket{\hat{h}}_{\beta=0}\right).
		\end{aligned}
	\end{equation}
	For sufficiently large system sizes, we expect both $E_0$ and $\braket{\hat{h}}_{\beta=0}$ to converge towards a finite value, and so in the infinite temperature limit where $Z=\mathcal{D}$ we recover $\braket{\hat{h}}_{\beta=0}$. On the other hand, at zero temperature we have that $Z=1$ and and we simply get $\tilde{E}_0=\braket{\hat{h}}_{\beta=\infty}$.
	
	If we are now interested in the deviation from the infinite temperature value, we find the simple expression
	\begin{equation}\label{eq:dev}
		\braket{\delta \hat{h}}(k\tau)=\frac{1}{Z}\left(\frac{\tilde{E}_0}{\braket{\hat{h}}_{\beta=0}}-1\right).
	\end{equation}
	In the simple case we consider here, we also have that $E_0=0$, leading to an even simpler formula: 
	\begin{equation}
		\braket{\delta\hat{h}}(k\tau)=-\frac{1}{Z}.
	\end{equation}
	
	If the revivals are not perfect, after one revival the ground state wavefunction does not lead to the value $\tilde{E}_0=0$.  Instead, it will go to a finite value that we denote by $h_k^\infty$, where the superscript $\infty$ denotes that this is the value at $\beta=\infty$ and the subscript $k$ that it is the value at the $t=k\tau$. In general, if scarring is not exact there no known analytical value for this quantity and we compute it numerically. So we simply replace $\tilde{E}_0$ by $h_k^\infty$ in the formula to get 
	\begin{equation}
		\braket{\delta \hat{h}}(k\tau)=\frac{1}{Z}\left(\frac{h_k^\infty}{\braket{\hat{h}}_{\beta=0}}-1\right).
	\end{equation}
	
	\section{Alternative preparation Hamiltonian for the XY model}\label{app:alternative}
	
	In order to compare more directly our results of the spin-1 XY model with those of the PXP model, here we use an alternative preparation Hamiltonian than holds a closer relation with the algebraic structure of the scarred states. This will have the effect of enhancing the overlap of states in the low-energy spectrum with scarred eigenstates.
	We now use the pre-quench Hamiltonian 
	\begin{equation} \label{eq:Hi_XY2}
		\hat{H}^{(2)}_\mathrm{i,XY}=\! \sum_{j=1}^N (-1)^j \left[\left(\hat{S}_j^x\right)^2\hspace{-0.2cm} {-}\left(\hat{S}_j^y\right)^2\right]
		\!{+}V\!\prod_{j=1}^N\left[\hat{\mathds{1}}_j{-} \left(\hat{S}_j^z\right)^2\right].
	\end{equation}
	Note that with respect to the preparation Hamiltonian in Eq.~\eqref{eq:Hi_XY}, the additional term $\propto V$ has been added. This has no effect on the ground state. However, if we choose $V\gg 1$ this heavily penalizes any occurrence of the $\ket{0}$ state. As a consequence, the first excited states have a single $(\ket{+1}\pm \ket{-1})/\sqrt{2}$ turned into $(\ket{+1}\mp \ket{-1})/\sqrt{2}$. The additional excitation will follow the same scheme, and the states with $\ket{0}$ sites --- which are orthogonal to scarred states --- will only contribute at large temperature. In practice, this means that the scarred subsapce will only have overlap on the ground state and on the $N$ first excited states, as is the case in the PXP model.
	
	One of the main consequences of the discussion above is that the symmetric superposition of the first set of excited states is entirely contained in the scarred subspace. Thus, one state among the $N$ first-excited states plays a role in the revivals. This is similar to the situation in the PXP model, where one state out of the $N/2$ in the first set of excitations belongs to the scarred subspace. In comparison, for the simpler $\hat{H}_\mathrm{i}$ in Eq.~\ref{eq:Hi_XY}, only every other set of low-energy excited states has a non-zero overlap on the scarred subspace, meaning that the first set of excited states is orthogonal to it. In addition, even when this overlap is non-zero (like for the second set of excited states), it is smaller as there are many states with $\ket{0}$ that are thus orthogonal to the scarred eigenstates.
	However, in all cases the contribution of the scarred subspace in excited states still vanishes as $N\to \infty$. 
	
	In the rest of this section, we set $V=50$. In order to adapt our analytic expectation to this change, we change $Z$ to
	\begin{align}\nonumber
		Z=&\,(1{+}e^{{-}2\beta})^N\\
		&{+}e^{{-}\beta V}\left[(1{+}e^{{-}\beta}{+}e^{-2\beta})^N{-}(1{+}e^{{-}2\beta})^N\right],
	\end{align}
	which is very close to simply $(1{+}e^{{-}2\beta})^N$ for $\beta>10^{-1}$.
	Indeed, while away from the $\beta \ll 1$ regime, the initial state has essentially no overlap with any state with a $\ket{0}$ site.
	
\begin{figure}[t!]
		\centering
		\includegraphics[width=\linewidth]{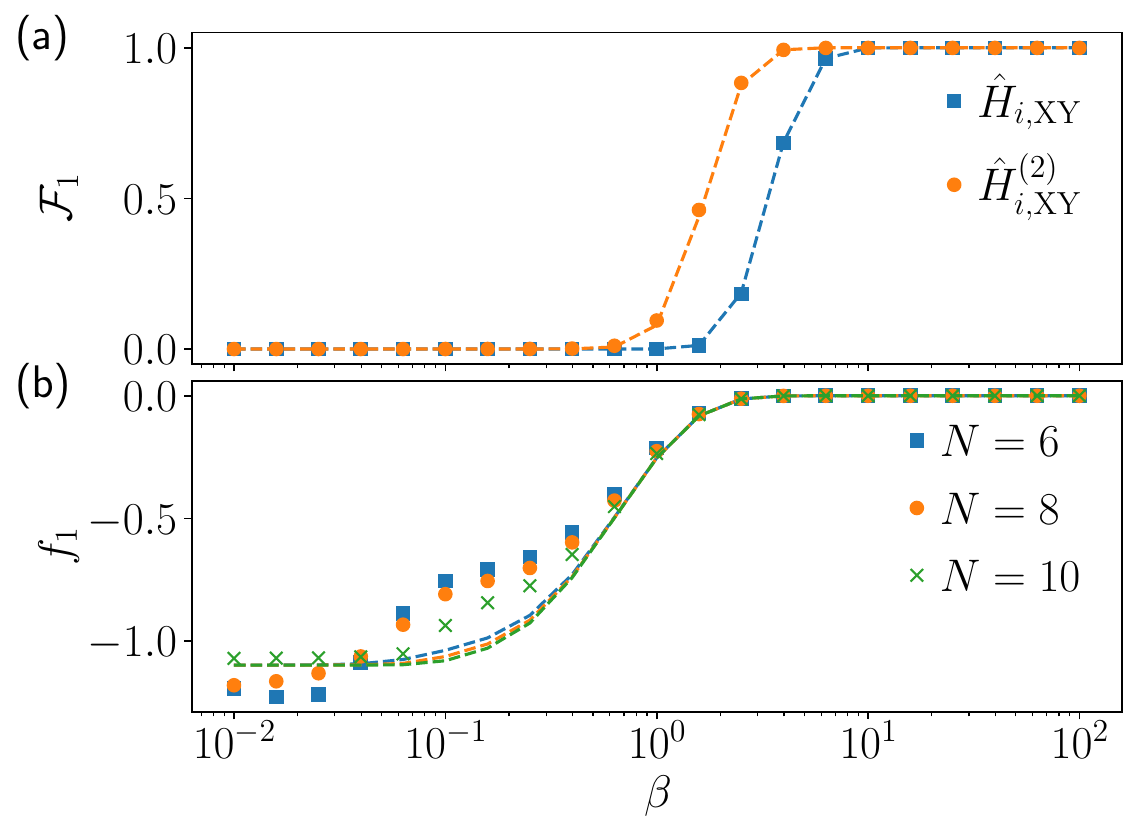}
		\caption{\small Maximum interferometric amplitude after a finite-$T$ quench in the spin-1 XY model for various values of inverse temperature. The dashed lines indicate the expected scaling for the data of the same color. (a) Comparison between the preparation Hamiltonians in Eqs.~\eqref{eq:Hi_XY} and \eqref{eq:Hi_XY2} for $N=10$. (b) Quenches using the preparation Hamiltonian in Eq.~\eqref{eq:Hi_XY2} for various system sizes. While there are some deviations from the expected behavior, they decay with system size and only happen for very low values of the fidelity density. }\label{fig:XY_fid}
	\end{figure}

\begin{figure}[tb]
		\centering
		\includegraphics[width=\linewidth]{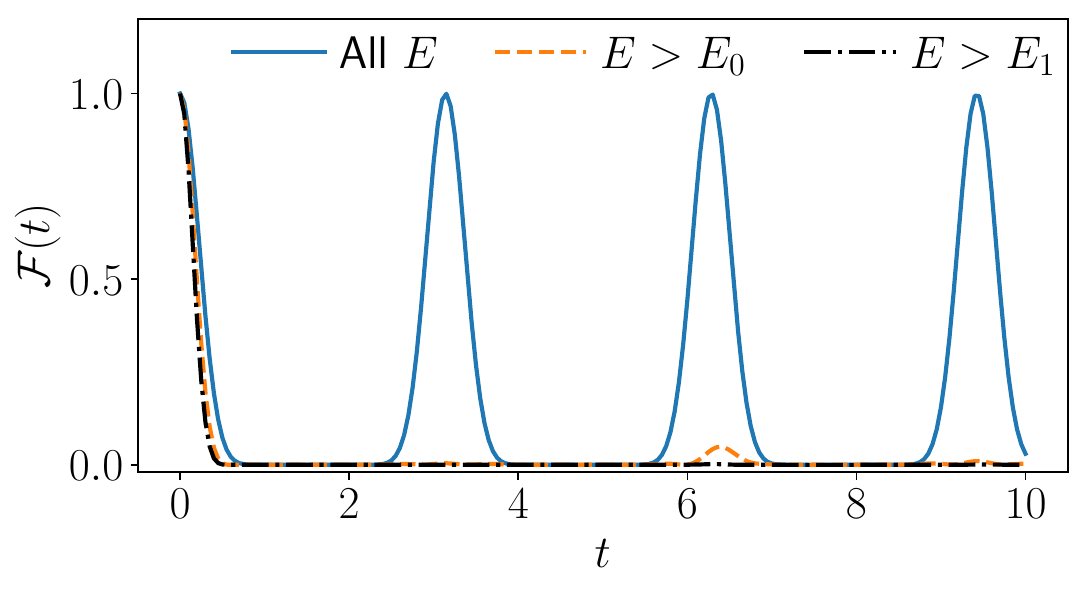}
		\caption{\small Interferometric amplitude after a finite-$T$ quench in the spin-1 XY model with $N=10$ using the preparation Hamiltonian in Eq.~\eqref{eq:Hi_XY2}. Various energy penalties on the low-energy spectrum are compared. The only visible difference with Fig.~\ref{fig:XY_fid_NVGS} is the slightly larger amplitude of the fluctuations around $t=2\tau$.}\label{fig:XY_no_GS}
	\end{figure}

	Results for quenches using the preparation Hamiltonian in Eq.~(\ref{eq:Hi_XY2}) are shown on Fig.~\ref{fig:XY_fid}.
	While we see a good fidelity compared to the original pre-quench Hamiltonian in Eq.~\eqref{eq:Hi_XY}, this is  due to the difference of the weight on the ground state as captured by the analytic prediction. There are also some small deviations with respect to the theoretical prediction for $\beta\approx 0.1$, but they decay rapidly with system size. They also happen in a regime where the observed fidelity is effectively zero, meaning that traces of scarring in the system will be extremely difficult to measure. This showcases that, as seen with the previous preparation Hamiltonian, only the ground state of $\hat{H}_\mathrm{i}$ is expected to contribute to the non-ergodic dynamics in the thermodynamic limit. This is confirmed by quenches where the contribution of the ground state is artificially removed by setting an energy penalty on it; see Fig.~\ref{fig:XY_no_GS}. While the peak in the middle panel is slightly larger than in Fig.~\ref{fig:XY_fid_NVGS}, the difference is small and expected to decay with system size.
	
	\section{Perturbed PXP model}\label{app:PXPpert}
	
	While the QMBS phenomenology in the spin-$1$ XY model resemble that of the PXP model discussed in the main text, one obvious difference is that the former hosts \emph{exact} QMBS and \emph{perfect} revivals. Thus, in order to be able to compare the two models on the same footing, we consider the perturbed version of the PXP model, $\hat{H}_\mathrm{f,PXP} + \delta \hat{H}$, in which scarring is essentially perfect. This perturbation was devised in Ref.~\cite{Choi2018} and takes the form
	\begin{equation}\label{eq:PXP_pert}
		\delta \hat{H} =\; -\sum_{j=1}^N \sum_{d=2}^{N/2}h_d \hat{P}_{j-1}\hat{\sigma}^x_j \hat{P}_{j+1}\left(\hat{\sigma}^z_{j-d}+\hat{\sigma}^z_{j+d} \right),
	\end{equation}
	with 
	\begin{equation}
		h_d=h_0\left(\phi^{d-1}-\phi^{1-d}\right)^{-2},
	\end{equation}
	$h_0=0.051$, and $\phi=(1+\sqrt{5})/2$ the golden ratio. The first order term in this expansion was also considered in Ref.~\cite{Khemani2018}. Low-order terms of an expansion such as Eq.~\eqref{eq:PXP_pert} can be iteratively derived in a process of ``correcting'' the structure constants of the $\mathrm{su}(2)$ algebra representation, furnished by QMBS eigenstates~\cite{Bull2020}. Thus, the perturbation in Eq.~\eqref{eq:PXP_pert} makes the revivals from the N\'eel state essentially perfect and the associated algebra in the QMBS subspace nearly $\mathrm{su}(2)$, allowing for a much closer comparison with the spin-$1$ XY model.

	\begin{figure}[t]
		\centering
		\includegraphics[width=\linewidth]{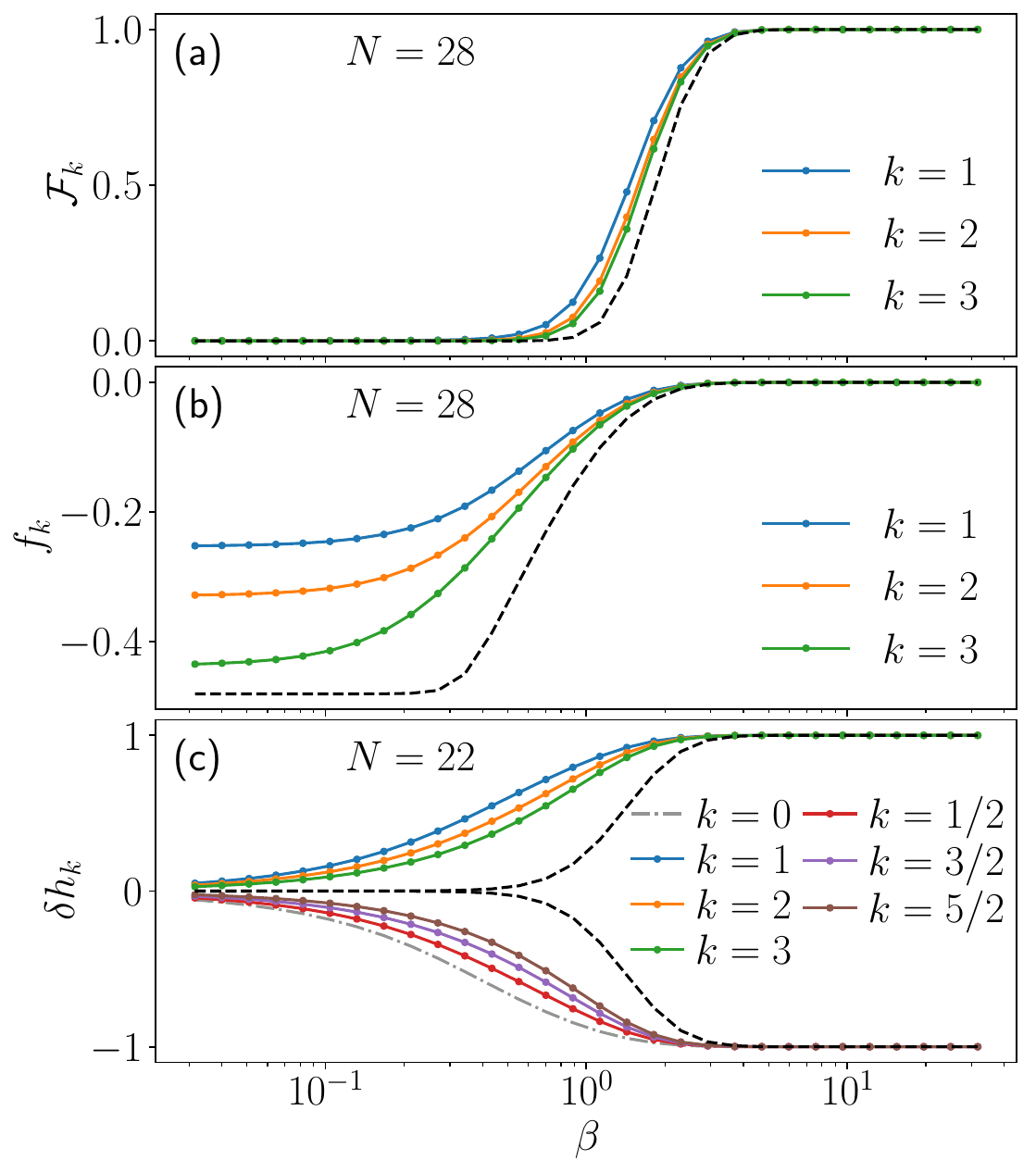}
		\caption{\small Fidelity, fidelity density, and deviation of observable density for various values of the temperature in the perturbed PXP model. All quantities show strong deviation from the naive expectation denoted by the dashed black lines. }\label{fig:PXP_pert_fid}
	\end{figure}
	
\begin{figure}[tbh]
		\centering
		\includegraphics[width=\linewidth]{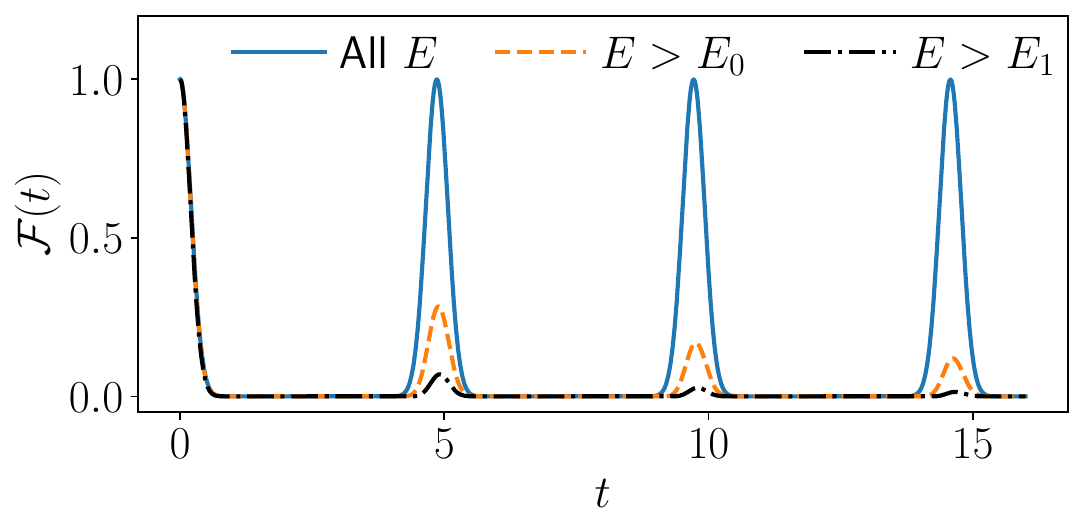}
		\caption{\small Interferometric amplitude after a quench in the perturbed PXP model for $N=28$ with energy penalties placed on the low-energy spectrum. Clear revivals can be seen even when the ground state and the first set of excited states are removed.}\label{fig:PXP_pert_no_GS}
	\end{figure}

	Using the perturbed PXP model in Eq.~\eqref{eq:PXP_pert}, we repeat the computations for the pure PXP model given in the main text, in order to check to what extent the exactness of QMBS structure impacts our conclusions. The 
	dynamics of $\mathcal{F}_k$, $f_k$, and $h_k$ for the perturbed PXP model are shown in Fig.~\ref{fig:PXP_pert_fid}.  For all metrics, we see strong deviations from the naive thermal predictions. 
	In Fig.~\ref{fig:PXP_pert_no_GS} we compute the fidelity in the case where the ground state is artificially brought to infinite energy. Not only are clear revivals visible when the ground state is excluded, the same is true when the first set of excitations is excluded as well. The symmetric superposition of all states with one defect on top of the N\'eel state should have overlap exclusively on scarred eigenstates. However, the other $N/2-1$ superpositions will be orthogonal to it, and should theoretically not contribute to the revivals. Thus, as only one state out of $N/2$ contribute, we expect its contribution to be similar to what was seen in the XY model. The next set of excitations is then made of the N\'eel state with two defects. As once again only the symmetric superposition is in the scarred subspace, this concerns one state in $N(N/2-1)/2$. Overall, one would expect the behavior to be the same as in the XY model, but it clearly is not.

\begin{figure}[t]
		\centering
		\includegraphics[width=\linewidth]{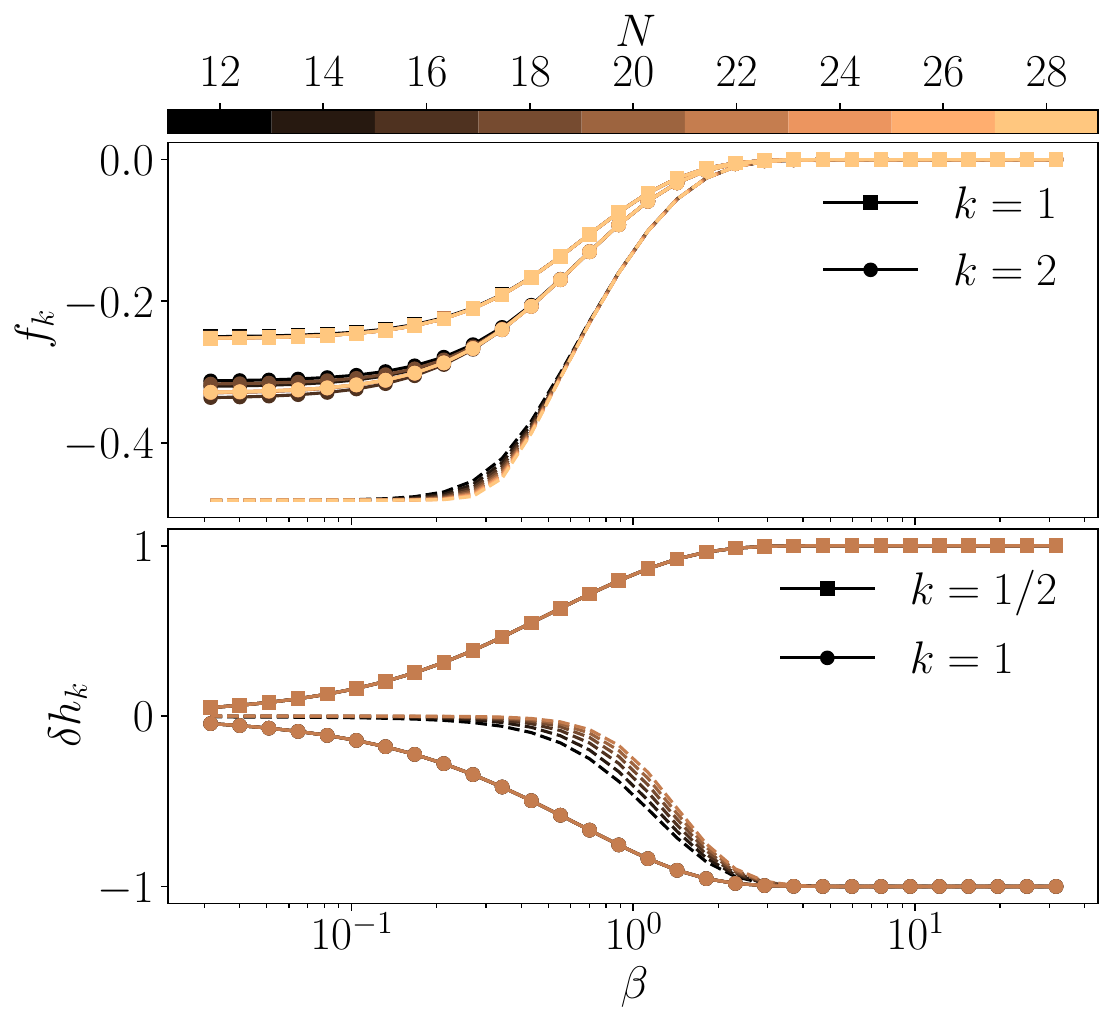}
		\caption{\small Scaling of fidelity density and deviation of $\hat{H}_\mathrm{i}/N$ after a quench in the perturbed PXP model. The dashed lines correspond to the theoretical expectations. Both metrics are well converged in system size and show robustness to finite temperature when compared to the expected behavior.}\label{fig:PXP_pert_scaling}
	\end{figure}

	We emphasize that what we witness in the PXP model is not a finite-size effect. Actually, the Hilbert space sizes explored in this model are \emph{larger} than the ones in the XY model. Indeed, in the latter we saw good agreement with the theoretical predictions already for $N=8$ and $N=10$, corresponding to $\mathcal{D}=3^{9}=19683$ and $\mathcal{D}=3^{10}=59049$, respectively.
	Meanwhile, in the PXP model we still see strong deviations in $f_k$ and $\mathcal{F}_k$ for $N=28$ where $\mathcal{D}=710647$, an order of magnitude larger. For $h_k$ we probed system sizes up to $N=22$ where $\mathcal{D}=39603$. Furthermore, we provide the scaling of $f_k$ and $\delta h_k$ with system size in Fig.~\ref{fig:PXP_pert_scaling}. Our results show that both quantities are well converged already at $N\approx 20$. As such, we expect the same special behavior in larger systems. This includes the higher-than-expected fidelity density near infinite temperature.
	
	\section{Details of the quantum algorithm}\label{app:algo}
	
	To simulate the PXP model, we have employed the IBM quantum processor, Kolkata, which uses a heavy hex topology and has quantum volume $128$ \cite{IBMdata}. The IBM processors use a cross-resonance gate to generate the CNOT entangling operation. On this hardware, we simulated the time dependence of the staggered magnetization, $\hat{M}_S$ in Eq.~\eqref{eq:H_prep1}. We simulate the evolution of the system under the Hamiltonian in Eq.~\eqref{eq:PXPHam} but now, for convenience, assuming open boundary conditions. The boundary terms in the Hamiltonian are taken to be  $\hat{\sigma}^x_1\hat{P}_2$ and $\hat{P}_{N-1}\hat{\sigma}^x_N$.
	
	As in all the classical simulations in this work, our goal is to simulate evolution for an initial Gibbs state at temperature $\beta$, as defined in Eq.~\eqref{eq:rho}. This must be done in the constrained Hilbert space where there are no neighboring $\ket{\uparrow}$, thus we only consider states in this subspace for our initial state. 
	The time dependence of $\hat{M}_S$ can be explicitly written as
	\begin{align}
		\langle \hat{M}_\mathrm{S}(t)\rangle {=} \sum_{E_k} \frac{e^{-\beta E_k}}{Z}\langle E_k | \hat{U}^{\dagger}(t) \hat{M}_\mathrm{S}\hat{U}(t)|E_k\rangle,
	\end{align}
	where we recall that the $\ket{E_k}$ are the eigenstates of $\hat{H}_\mathrm{i}$ in the constrained Hilbert space.
	At this point, we can see that it is sufficient to perform the simulation for all states in $\hat{\rho}$, and perform a weighted average using their Boltzmann weights, $e^{-\beta E_k}/Z$.
	
	We prepare the thermal state from Eq.~\eqref{eq:rho} using the $E\rho Oq$ method \cite{Gustafson:2020yfe,Lamm:2018siq,Harmalkar:2020mpd}.
	This method involves sampling states from the density matrix using traditional Markov Chain Monte Carlo (MCM) methods rather than preparing the thermal state explicitly on the quantum computer.
	
	We generate configurations from the Hamiltonian in Eq.~\eqref{eq:H_prep1} as follows. 
	Because that Hamiltonian is diagonal, the density matrix can be written as a diagonal operator 
	\begin{equation}
		\hat{\rho}_{i}(\beta) = \frac{1}{Z}\sum_{\lbrace\mathcal{S}_j\rbrace} e^{-\beta E_{\mathcal{S}_j}}|\mathcal{S}_j\rangle\langle \mathcal{S}_j|,
	\end{equation}
	where the sum over $\lbrace\mathcal{S}_j\rbrace$ includes only the allowed spin configurations.
	We can now identify a corresponding action $S = \hat{H}_\mathrm{i}$. 
	The system is prepared in a valid spin configuration and spin changes are proposed randomly in MCMC sweeps with a given probability weighted by the change in total energy: $e^{-\beta (E_{\mathcal{S}'} - E_{\mathcal{S}})}$.
	If a proposed change would take the system to an invalid subspace the proposed change is discarded.
	After generating $N_c$ configurations of the form $|\mathcal{S}_j\rangle\langle\mathcal{S}_j|$ from the density matrix, we then simulate the time dependence of $\hat{M}_S$ for each unique spin configuration.
	The thermal average is then the weighted average, 
	\begin{align}
		\langle \hat{M}_\mathrm{S}(t)\rangle = \sum_{\lbrace\mathcal{S}_i\rbrace} \frac{p_i}{N_c} \langle \mathcal{S}_i | \hat{U}^{\dagger}(t) \hat{M}_\mathrm{S}\hat{U}(t)|\mathcal{S}_i\rangle, 
	\end{align}
	where $p_i$ is the number of times the configuration appeared in the simulation. 
	If an nondiagonal Hamiltonian is used for state preparation then linear combinations of the bra and ket vectors in the density matrix need to be used.
	In principle the accuracy of this method encounters an exponential signal to noise problem that is slightly lessened by the use of a diagonal Hamiltonian \cite{Harmalkar:2020mpd,Lamm2020}.

	We used the suite of error mitigation techniques provided by QISKit Runtime \cite{qiskit2024}, which include: dynamic decoupling \cite{Ezzell2022,Qi:2022gdn,Morong2023, Jurcevic_2021,Niu:2022wpa,Niu:2022jnx,Mundada:2022roq,qiskit2024}, randomized compiling \cite{2016efficienttwirling,Erhard_2019,li2017efficient, 2018efficienttwirling, 2013PhRvA..88a2314G, 2016efficienttwirling,Silva-PT2008,Winick:2022scr}, and readout mitigation (specifically T-REx) \cite{PhysRevApplied.14.054059,PhysRevApplied.12.054023,PhysRevApplied.10.034040,Sarovar2020detectingcrosstalk,Berg:2020ibi,Smith:2021iwt,Rudinger:2021nhd,10.1145/3352460.3358265,Harrigan2021,PhysRevA.100.052315,Maciejewski2020mitigationofreadout,Nachman2020,Hicks:2021uvm,PhysRevLett.122.110501,PhysRevA.101.032343,Hamilton:2020xpx,Geller_2021,PhysRevLett.119.180511}. 
	Dynamic decoupling is a method which aims to tackle dephasing errors that a quantum state accumulates by frequent applications of quantum gates on idling qubits which act to cancel accumulated phase errors. 
	Randomized compiling is used to transform the unitary errors from the CNOT gate being imperfect into random stochastic Pauli errors which are typically less catastrophic. 
	Readout mitigation is a tool which takes the output probability distribution measured from the quantum computer and changes the relative bitstring outputs using apriori knowledge determined when the quantum computer is calibrated on the likelihood of misidentifying a $|0\rangle$ or $|1\rangle$ state. 
	We also used a rescaling procedure to counteract the signal loss from the effective depolarizing channel caused by the randomized compiling \cite{Urbanek2021,Vovrosh:2021ocf,ARahman:2022tkr}.
	This method works by running a circuit which contains only Clifford gates and has a known classical output and using the discrepancy between the measured and expected value to renormalize the observed value.
	
	\begin{figure*}[htb]
		\begin{equation*}
			e^{i \delta t P_j X_{j + 1} P_{j + 2}} =
			\begin{gathered}
				\Qcircuit @R=1em @C=1em {
					& \qw & \ctrl{1} & \qw & \qw & \qw & \qw & \ctrl{1} & \qw & \qw & \qw & \qw & \qw & \qw \\
					& \gate{H} & \targ{} & \gate{R_z(\delta t)} & \ctrl{1} & \qw & \ctrl{1} & \targ & \ctrl{1} & \qw & \ctrl{1} & \gate{R_z(-\delta t)} & \gate{H} & \qw \\
					& \qw & \qw & \qw & \targ & \gate{R_z(-\delta t)} & \targ & \qw & \targ & \gate{R_z(\delta t)} & \targ & \qw & \qw & \qw 
				}
			\end{gathered}
		\end{equation*}
		\caption{\small Quantum circuit that implements the PXP operation.}
		\label{fig:pxpoperator}
	\end{figure*}
	To help with reproducibility we include the circuit decomposition in terms of CNOTs and native two qubit gates in Fig.~\ref{fig:pxpoperator}. This operator is then tessellated across the lattice of spins using a Trotter decomposition where the operator is decomposed as
	\begin{eqnarray}
		\notag   \hat{U}(\delta t) &=& e^{i\delta t \sum_{i=0} P_{3i}X_{3i+1}P_{3i+2}} \\ 
		\notag     && e^{i\delta t \sum_{i=0} P_{3i+1}X_{3i+2}P_{3i+3}} \\
		&& e^{i\delta t \sum_{i=0} P_{3i+2}X_{3i+3}P_{3i+4}}.    
	\end{eqnarray}

	\bibliography{biblio}
	
	\clearpage

\end{document}